\documentclass[aps,prd,twocolumn,superscriptaddress,groupedaddress,nofootinbib]{revtex4}
\usepackage{graphicx}  
\usepackage{dcolumn}   
\usepackage{bm}        
\usepackage{amssymb}   
\usepackage{amsmath}
\usepackage{dsfont}
\usepackage{xcolor}
\usepackage{amsthm}
\usepackage{enumitem}
\usepackage{tensor}
\usepackage[colorlinks]{hyperref}
\hypersetup{
     breaklinks=true,
    pdfstartview={FitH},    
    colorlinks=true,       
    linkcolor=blue,          
    citecolor=red,        
    filecolor=magenta,      
    urlcolor=blue,           
    anchorcolor=green,      
    linktocpage=true
}

\newcommand{\be}{\begin{equation}}\newcommand{\ee}{\end{equation}}
\newcommand{\bea}{\begin{eqnarray}}\newcommand{\eea}{\end{eqnarray}}
\newcommand{\brr}{\begin{array}}\newcommand{\err}{\end{array}}
\newcommand{\bit}{\begin{itemize}}\newcommand{\eit}{\end{itemize}}
\newcommand{\ben}{\begin{enumerate}}\newcommand{\een}{\end{enumerate}}

\newcommand{\ba}{\begin{array}}
\newcommand{\ea}{\end{array}}

\definecolor{darkred}{rgb}{.8,0,0}
\newcommand{\cdred}{\color{darkred}}

\definecolor{darkblue}{rgb}{0,0,0}
\newcommand{\cdblue}{\color{darkblue}}

\def\1{{_{1}}}\def\2{{_{2}}}

\def\noHe0{:\;\!\!\;\!\!:H_e(0):\;\!\!\;\!\!:}
\def\noHm0{:\;\!\!\;\!\!:H_\mu(0):\;\!\!\;\!\!:}
\begin{document}

\title{Newman--Penrose formalism and exact vacuum solutions to conformal Weyl gravity}
\author{Petr Jizba}
\email{p.jizba@fjfi.cvut.cz}
\affiliation{FNSPE,
Czech Technical University in Prague, B\v{r}ehov\'{a} 7, 115 19, Prague, Czech Republic}
\author{Kamil Mudru\v{n}ka}
\email{mudrukam@fjfi.cvut.cz}
\affiliation{FNSPE,
Czech Technical University in Prague, B\v{r}ehov\'{a} 7, 115 19, Prague, Czech Republic}
\date{\today}
\begin{abstract}
We study the exterior solution for a static, spherically symmetric source in Weyl conformal gravity in terms of the Newman--Penrose formalism. We first show that both the static, uncharged black hole solution of Mannheim and Kazanas and the static, charged Reissner--Nordstr\"{o}m-like solution can be found more easily in this formalism than in the traditional coordinate-basis approach, where the metric tensor components are taken as the basic variables. 
Second, we show that the Newman-Penrose formalism offers a particularly convenient framework that is well suited for the discussion of conformal gravity solutions corresponding to Petrov ``type-D'' spacetimes.
This is illustrated with a two-parameter class of wormhole solutions that  
includes the  Ellis--Bronnikov wormhole solution of Einstein's gravity as a limiting case.
Other salient issues, such as the gauge equivalence of solutions and the inclusion of the electromagnetic field are also discussed.

\end{abstract}

\vskip -1.0 truecm

\maketitle

\section{Introduction}
\label{Intro}

Currently, the dominant classical theory of gravity remains Einstein's general theory of relativity, which has been remarkably successful in explaining gravitational experiment data at solar-system scales and smaller. 
The viability of this approach at larger scales, such as the scales of galaxies, galaxy clusters, superclusters, galactic walls, etc., is less clear, since a significant body of evidence suggests a discrepancy between the properties of the Universe as predicted by general relativity (coupled to known matter) and what is actually observed~\cite{Riess,Perlmutter,Spergel1,Spergel2,Tegmark,Eisenstein}. 
The effects of this mismatch can be incorporated into the framework of general relativity by introducing additional matter described as a nearly-pressureless, perfect fluid --- that is, dark matter ---but it is unclear whether this represents the effect of a truly new exotic matter field or a manifestation of the breakdown of Einstein's general relativity. On the other hand, one can generalize Einstein's theory to some higher-derivative gravity. Higher-derivative gravitational theories are important contenders for a viable gravity theory, both because they naturally allow one to avoid Lovelock's no-go theorem~\cite{Lov:71,Lov:72} (which applies only to second-derivative gravity theories), and because they can often be quantized (or are at least perturbatively renormalizable). A particularly important higher-derivative gravity theory is Weyl conformal gravity (WCG), which has interesting classical phenomenology and elegantly addresses the dark matter problem~\cite{mannheim_b,mannheim_c,MK2,Kazanas,mannheim,mannheim_a,JKS2,Edery:1998zi,Edery:1997hu,Edery:2001at}.

In this paper we present a derivation of the exterior solution for a static, spherically symmetric source in WCG. In contrast to more traditional approaches, such as the coordinate-basis approach using the metric tensor components as the basic variable and the Christoffel symbols for the  connection~\cite{MK2,Kazanas,mannheim,mannheim_a}, or the approach based on Cartan's calculus of differential forms~\cite{Charmousis,Attard}, we employ here the Newman--Penrose (NP) formalism, i.e. a technique based on the use of null tetrads, with ideas taken from 2-component spinors~\cite{NP1,NP2}. This not only provides a quick path to such conventional solutions as the static, uncharged black hole solution of Mannheim and Kazanas and the static, charged Reissner--Nordstr\"{o}m type solution, but also allows us to tackle new, more challenging problems such as the wormhole solutions.

The standard treatment of problems in higher-derivative theories of gravity (including WCG) is based on considering the field equations (for WCG, the so-called Bach equation) in a local coordinate basis adapted to the problem at hand. However, this may not be the only strategy. For example, in some contexts in Einstein's general theory of relativity, it has been found convenient to use a different approach, namely to choose a suitable tetrad basis of four linearly independent vector fields, to project the relevant quantities onto the chosen basis, and to consider the equations satisfied by them~\cite{Chandra}.
%
%
In the tetrad formalism, the choice of the tetrad basis depends on the underlying symmetries of the space-time one wishes to capture. A special choice of tetrad basis vectors, consisting of a tetrad of zero vectors, is particularly suitable for conformal gravity, where there is no mass scale. Such a choice of the tetrad basis constitutes the Newman--Penrose formalism~\cite{NP1,NP2}. Moreover, one might expect that for black hole solutions in WCG, the NP formalism will enjoy the same status as in Einstein's general relativity, where the equations of motion are substantially simplified due to the Goldberg--Sachs theorem~\cite{GS}, which implies,  among other things, that the resulting black hole solutions correspond to spacetimes of Petrov ``type-D'' character.  We will, indeed, confirm the latter also in the case of WCG.  


The paper is organized as follows: in the following section we briefly revisit some of the aspects of Weyl conformal gravity, which will be required in the main body of the text. The role of the Newman--Penrose formalism in  WCG is outlined in Sec.~\ref{Sec.2}. In Sec.~\ref{Sec.3} we demonstrate the effectiveness of the NP formalism in WCG by re-deriving quickly the Mannheim--Kazanas and Reissner--Nordstr\"{o}m type solutions.   In Sec.~\ref{Sec.V.cd} we outline a general framework for solving the vacuum Bach equations in the NP formalism and in arbitrary gauge. Our particular focus will be on Petrov ``type-D'' spacetimes, which will considerably simplify the analysis.
Besides the already obtained solutions, new wormhole solutions with and without electromagnetic field will be derived and discussed. {\cdblue Using both geometric and topological arguments, we also show that the obtained wormhole solutions are not Weyl equivalent to the WCG black hole solutions.}
A brief summary of results and related discussions are provided in Sec.~\ref{Concl.}.
For the reader's convenience, we relegate some more technical issues concerning the Bach equation in the NP formalism to two appendices.

\section{Weyl conformal gravity}\label{Sec.I}

For the sake of consistency, we will now briefly review the key aspects of the WCG that will be needed in the following sections. More detailed exposition can be found, e.g. in Ref.~\cite{RJ}.

{WCG is a pure metric theory that is invariant not only under the action of
the diffeomorphism group, but also under the Weyl rescaling of the metric tensor by 
non-singular smooth functions $\Omega(x)$: $g_{\mu\nu}(x)\rightarrow \Omega^2(x)g_{\mu\nu}(x)$.
The simplest WCG action functional in four spacetime dimensions has the form~\cite{Weyl1,Bach},
\begin{equation}
S[g] \ = \ -\frac{1}{4G_{\rm{w}}^{\tiny{2}}}\int d^4x \ \! \sqrt{|g|} \ \! C_{\mu\nu\rho\sigma}
C^{\mu\nu\rho\sigma}\, .
\label{PA1}
\end{equation}
Here  $C_{\mu\nu\rho\sigma}$ is the
{Weyl tensor} which can be written as
\begin{eqnarray}
C_{\mu\nu\rho\sigma}  &=&
R_{\mu\nu\rho\sigma} \ - \ \left(g_{\mu[\rho}R_{\sigma]\nu} -
g_{\nu[\rho}R_{\sigma]\mu} \right)
\nonumber\\[1mm]
&+& \frac{1}{3}R \ \! g_{\mu[\rho}g_{\sigma]\nu}\, ,
\label{P2a}
\end{eqnarray}
with $R_{\mu\nu\rho\sigma}$ being the {Riemann curvature tensor},
$R_{\mu\rho}=R_{\mu\nu\rho}{}^\nu$  the  {Ricci tensor}, and $R= R_\mu{}^\mu$
the {scalar curvature}.  The Weyl tensor has all the algebraic properties
of the Riemann curvature tensor, i.e. 
\begin{eqnarray}
&&C_{\alpha \beta \gamma \delta} \ = \  C_{[\alpha \beta][\gamma \delta]} \ = \  C_{\gamma \delta \alpha \beta}\, ,\nonumber \\[2mm] 
&&C_{\alpha[\beta\gamma\delta]} \ = \ 0\, ,
\end{eqnarray}
and in addition is trace-free, i.e. $C_{\alpha \beta \gamma}^{~~~~\alpha} = 0$. 
Weyl tensor is identically zero in three dimensions.

Here and throughout, we employ the space-like metric signature $(-,+,+,+)$. 
The dimensionless coupling constant $G_{\rm{w}}$ is chosen to mimic the Yang--Mills action. 
As for the notation for various scalar invariants (with four derivatives of the metric tensor), we accept the following conventions: for the square of
the Riemann tensor contracted naturally (preserved order of indices), that is $R_{\mu\nu\rho\sigma}R^{\mu\nu\rho\sigma}$, we use the symbol
$R^2_{\mu\nu\rho\sigma}$; the square of the Ricci tensor $R_{\mu\nu}R^{\mu\nu}$, we denote by simply $R_{\mu\nu}^2$; the square of the Ricci
curvature scalar is always $R^2$, while for the Weyl tensor square (with a natural contraction of indices) $C_{\mu\nu\rho\sigma}C^{\mu\nu\rho\sigma}$,
we employ a shorthand and schematic notation $C^2$. When the latter is treated as a local invariant (not under a volume integral, so without the possibility of
integrating by parts) in $d=4$ dimensions, one finds the following expansion of the $C^2$ invariant into standard invariants quadratic in curvature
\begin{equation}
C^2 \ = \ R_{\mu\nu\rho\sigma}^2 \ - \ 2R_{\mu\nu}^2 \ + \ \frac{1}{3}R^2\,.
\label{weylsquare}
\end{equation}
Finally, we can employ yet another important combination of the quadratic curvature invariants, namely Gauss--Bonnet (GB) term
\begin{equation}
E_4 \ = \ R_{\mu\nu\rho\sigma}^2 \ - \ 4R_{\mu\nu}^2 \ + \ R^2\, ,
\label{GBdef}
\end{equation}
which in $d=4$ is the integrand of the Euler–Poincaré invariant~\cite{Percacci}
\begin{eqnarray}
\chi \ = \ \frac{1}{32\pi^2}\int d^4x \ \! \sqrt{|g|} \ \! E_4\, .
\label{II.8.aa}
\end{eqnarray}
With the help of the Chern--Gauss--Bonnet  theorem, one
can cast the Weyl action $S$ into an equivalent form (modulo topological term)
\begin{equation}
S[g] \ = \
-\frac1{2G_{\rm{w}}^2}\int d^4x\ \! \sqrt{|g|} \ \!\left(R_{\mu\nu}^2 \ - \ 
\frac{1}{3} R^2\right).
\label{PA2}
\end{equation}
It should be stressed that both (\ref{PA1}) and (\ref{PA2}) are Weyl-invariant only in $d=4$ dimensions.
In fact, under the Weyl transformation $g_{\mu \nu} \to \Omega^{2} {g}_{\mu \nu}$, the densitized square of the Weyl tensor
transforms as
\begin{eqnarray}
\sqrt{|g|} \ \! C^2 \ \to \ \Omega^{d-4} \sqrt{|g|}   C^2\, ,
\end{eqnarray}
in a general  $d$-dimensional spacetime. At the same time, the term $\sqrt{|{g}|} \ \!E_4$  supplies topological (and thus also Weyl) invariant only in $d=4$.} 
It should be emphasized that although the topological term is clearly not important at the classical level, it is relevant at the quantum level, where summation over different topologies should be considered. However, even if one stays on topologies with a fixed Euler--Poincar\'{e} invariant, the renormalization procedure will inevitably generate (already at a one loop) the GB term~\cite{Fradkin:1985am}.  
Variation of $S$ with respect to the metric
yields the field equation --- {\em Bach vacuum equation}~\cite{Bach}
\begin{equation}
\nabla^{\kappa} \nabla^{\lambda} C_{\mu\kappa\nu\lambda
}
\ - \ \frac{1}{2} \ \! R^{\kappa\lambda} C_{\mu\kappa\nu\lambda} \ \equiv \ B_{\mu\nu}
\ = \  0\, ,
\label{Z42A}
\end{equation}
where $B_{\mu\nu}$  is the Bach tensor and  $\nabla$
is the Levi-Civita connection, i.e. the connection that is metric compatible,
and torsion-free. It is easy to see that $B_{\mu\nu}$ is symmetric and trace-free. 
Backgrounds satisfying~(\ref{Z42A}) are known as {\em Bach-flat} backgrounds.

Making use of the contracted Bianchi
identity (i.e. traces of the usual second Bianchi identity of the Riemann tensor) in the form 
\begin{eqnarray}
\nabla^{\delta} C_{\alpha \beta \gamma \delta } \ = \ - \nabla_{[\alpha} \!\left(R_{\beta]\gamma} \ - \ \frac{1}{6} R\ \! g_{\beta]\gamma}\right), 
\end{eqnarray}
we might, in passing, note that the Bach tensor can be rewritten as
\begin{eqnarray}
B_{\mu \nu} &=& 2\nabla^{\kappa}\nabla_{[\kappa} P_{\mu]\nu} \ - \ P^{\kappa \lambda} C_{\mu\kappa\nu\lambda}\, ,
\label{II.11.cd}
\end{eqnarray}
where  $P$ is the Schouten tensor~\cite{Besse}, i.e.
\begin{eqnarray}
P_{\alpha\beta} \ = \ \frac{1}{2} \left(R_{\alpha \beta} \ - \ \frac{1}{6} \ \! R \ \! g_{\alpha\beta} \right).
\end{eqnarray}
Einstein's field equations for pure gravity with cosmological constant  correspond to $R_{\alpha \beta} = \mathbf{\Lambda} \ \! \!g_{\alpha \beta}$ (and hence $R=4 \mathbf{\Lambda}$). Spacetimes that satisfy these equations are known as {\em Einstein spaces}. Since for Einstein spaces $P_{\alpha\beta} = \mathbf{\Lambda} \ \! \! g_{\alpha\beta}/6$, we see from (\ref{II.11.cd}) that every Einstein space is also a solution of Bach's vacuum equation. The reverse is obviously not true. 

\vspace{3mm}
\section{Conformal gravity and Newman--Penrose formalism}
\label{Sec.2}

We begin with a brief overview of the NP formalism and its connection to WCG. This will also help us to set up the notation and sign conventions. The NP formalism is a tetrad formalism with a specific selection of the basis vectors~\cite{NP1,NP2}. The choice made is a tetrad of null vectors $\{l^{\mu}, n^{\mu}, m^{\mu}, \bar{m}^{\mu}\}$,  where $l^\mu$ and $n^\mu$ are real and $m^\mu$ and $\bar{m}^\mu$ are complex conjugates of each other. Penrose chose such a tetrad because he believed that a space-time's light-cone structure is essential for incorporating spinor calculus into general relativity.
It turns out that the special adaptability of the NP formalism to black hole solutions (which are of interest here) is due to their ``type-D'' character --- similarly as in Einstein's general relativity. This, in turn, will greatly simplify our subsequent analysis. 

Below, we accept the following normalization of tetrad null vectors
\begin{eqnarray}
\label{eq:NPTetradRelations}
&&l^{\mu}l_{\mu} \ = \ n^{\mu}n_{\mu} \ = \ m^{\mu}m_{\mu} \ = \ \bar{m}^{\mu}\bar{m}_{\mu} \ = \ 0\, , \nonumber \\[2mm]
&&l^{\mu}m_{\mu} \ = \ l^{\mu}\bar{m}_{\mu} \ = \ n^{\mu}m_{\mu} \ = \ n^{\mu}\bar{m}_{\mu} \ = \ 0\, , \nonumber \\[2mm]
&& l^{\mu}n_{\mu} \ = \ -1\, , \; \; m^{\mu}\bar{m}_{\mu} \ = \ 1\, .
\end{eqnarray}
In other words, the metric components in the local basis $\{l^{\mu}, n^{\mu}, m^{\mu}, \bar{m}^{\mu}\}$ are 
\begin{eqnarray}
g_{ab} \ = \ e^{\mu}_ae_{b \mu} \ = \  g^{ab} \ = \  \left(
   \begin{array}{cccc}
     0 & -1 & 0 & 0 \\
     -1 & 0 & 0 & 0 \\
     0 & 0 & 0 & 1 \\
     0 & 0 & 1 & 0 \\
   \end{array}
 \right)\, ,
\end{eqnarray}
where $e_1^\mu = l^\mu$, $e_2^\mu =  n^\mu$, $e_3^\mu = m^\mu$ and $e_4^\mu = \bar{m}^\mu$.

We denote the covariant derivatives in the direction of the tetrad vectors as $l^{\mu}\nabla_{\mu} = \mathrm{D}$, $n^{\mu}\nabla_{\mu} = \Delta$, $m^{\mu}\nabla_{\mu} = \delta$ and $\bar{m}^{\mu}\nabla_{\mu} = \bar{\delta}$.
Information about the geometry is then  contained in the spin (or Ricci rotation) coefficients i.e., the tetrad components of the covariant derivatives of the tetrad vectors
\begin{eqnarray}
   && \kappa \ = \  -m^{\mu}\mathrm{D} l_{\mu}\, , \;\;\;\; 
   \pi \ = \  \bar{m}^{\mu}\mathrm{D} n_{\mu}\, ,
    \nonumber \\[2mm] 
   &&\sigma \ = \  -m^{\mu}\delta l_{\mu}\, , \;\;\;\; 
   \mu \ = \  \bar{m}^{\mu}\delta n_{\mu}\, , 
   \nonumber \\[2mm]
   && \tau  \ = \  -m^{\mu}\Delta l_{\mu}\, , \;\;\;\; \nu \ = \  \bar{m}^{\mu}\Delta n_{\mu}\, , \nonumber \\[2mm] 
   && \rho \ = \  -m^{\mu}\bar{\delta} l_{\mu}\, ,\;\;\;\; \lambda \ = \  \bar{m}^{\mu}\bar{\delta} n_{\mu}\, , \nonumber \\[2mm]
    &&\alpha  \ = \  -\frac{1}{2}\left(n^{\mu}\bar{\delta} l_{\mu} \ - \ \bar{m}^{\mu}\bar{\delta} m_{\mu}\right)\, , \nonumber \\[2mm]
    &&\beta \ = \  -\frac{1}{2}\left(n^{\mu}\delta l_{\mu} \ - \ \bar{m}^{\mu}\delta m_{\mu}\right)\, , \nonumber \\[2mm]
    &&\gamma \ = \  -\frac{1}{2}\left(n^{\mu}\Delta l_{\mu} \ - \ \bar{m}^{\mu}\Delta m_{\mu}\right)\, , \nonumber \\[2mm] 
    &&\epsilon \ = \  -\frac{1}{2}\left(n^{\mu}\mathrm{D} l_{\mu} \ - \ \bar{m}^{\mu}\mathrm{D} m_{\mu}\right)\, ,
    \label{III.15.cc}
\end{eqnarray}
while the curvature is encoded in five complex NP-Weyl scalars
\begin{eqnarray}
    &&\mbox{\hspace{-4mm}}\Psi_0 \ = \ l^{\mu}m^{\nu}l^{\alpha}m^{\beta}\tensor{C}{_\mu_\nu_\alpha_\beta}\, , \;\;\;\;  \Psi_1 \ = \ l^{\mu}n^{\nu}l^{\alpha}m^{\beta}\tensor{C}{_\mu_\nu_\alpha_\beta}\, , \nonumber \\[2mm]
&&\mbox{\hspace{-4mm}}\Psi_2 \ = \ l^{\mu}m^{\nu}\bar{m}^{\alpha}n^{\beta}\tensor{C}{_\mu_\nu_\alpha_\beta}
    \, , \;\;\;\;
    \Psi_3 \ = \ l^{\mu}n^{\nu}\bar{m}^{\alpha}n^{\beta}\tensor{C}{_\mu_\nu_\alpha_\beta}\, ,\nonumber \\[2mm]
&&\mbox{\hspace{-4mm}}\Psi_4 \ = \ n^{\mu}\bar{m}^{\nu}n^{\alpha}\bar{m}^{\beta}\tensor{C}{_\mu_\nu_\alpha_\beta}\, ,
\label{III.16.cc}
\end{eqnarray}
3 complex NP-Ricci scalars 
\begin{eqnarray}
     \Phi_{01} &=& \frac{1}{2}R_{\mu\nu} l^{\mu}m^{\nu}\, , \;\;\;\; \Phi_{02} \ = \  \frac{1}{2}R_{\mu\nu} m^{\mu}m^{\nu}\, , \nonumber \\[2mm] \Phi_{12} &=& \frac{1}{2}R_{\mu\nu} n^{\mu}m^{\nu}\, ,
\end{eqnarray}
and 4 real NP-Ricci scalars
\begin{eqnarray}
    \Phi_{00} &=& \frac{1}{2}R_{\mu\nu} l^{\mu}l^{\nu}\, , \;\;\;\; \Phi_{22} \ = \  \frac{1}{2}R_{\mu\nu}n^{\mu}n^{\nu}\, , \nonumber \\[2mm]
    \Phi_{11} &=&  \frac{1}{4}\left(R_{\mu\nu} l^{\mu}n^{\nu} \ + \ R_{\mu\nu} m^{\mu}\bar{m}^{\nu}\right)\, , \nonumber \\[2mm] \Lambda &=&  \frac{R}{24}\, .
    \label{III.19.cv}
\end{eqnarray}
In particular, note that in the NP formalism the ten independent components of the Weyl tensor are represented by the five complex scalars, and the ten components of the Ricci tensor are defined in terms of the four real and three complex scalars.
Quantities (\ref{III.15.cc})-(\ref{III.19.cv}) are mutually related by the NP-equations. In Appendix~\ref{Appendix A0} we list the ensuing spin-coefficient equations.
The corresponding Bach tensor~(\ref{II.11.cd}) in the NP formalism is discussed in Appendix~\ref{Appendix A}.

{\cdblue

There are several clear advantages to using a tetrad based formalism over the metric based one in WCG. Let us now list some of them. {\em First}, since the Bach equations involve covariant derivatives of the Weyl tensor, and in general covariant derivatives of up to 4th order, the explicit evaluation of the terms involves a significant number of Christoffel symbols. On the other hand, the NP quantities are all scalars and therefore only partial derivatives are involved.

{\em Second}, the NP formalism is closely related to the Petrov classification. 
Unlike the Einstein field equations, which become algebraic constraints on the NP-Ricci scalars and thus do not provide a direct straightforward link to the Petrov classification, the Bach equations~\eqref{Z42A} represent a linear expression in the Weyl tensor components. As a result, using the NP formalism leads to a considerable simplification of the equations if only solutions of a fixed Petrov type are sought.

{\em Third},  various NP scalars transform conveniently simply under Weyl transformations. In particular, the transformation $\tensor{g}{_\mu_\nu}\rightarrow\Omega^2(x)\tensor{g}{_\mu_\nu}$ rescales the tetrad to
\begin{eqnarray}
    l_\mu \ \rightarrow \ \Omega l_\mu\, , \;\;\;\;  l^\mu \ \rightarrow \ \Omega^{-1} l^\mu\, ,~~~~~ 
\end{eqnarray}
and the same rule applies to the other three base vectors. They are kept null and the normalization relations~\eqref{eq:NPTetradRelations} are preserved. The NP-Weyl scalars transform trivially as $\Psi_i\rightarrow\Omega^{-2}\Psi_i$. Thus, we see that the Petrov types (formulated in terms of Weyl scalars) are preserved by Weyl transformations, and the restrictions of the Bach equations on individual types are therefore Weyl invariant. Transformations for the NP-Ricci scalars can be found by taking projections of the relation
\begin{eqnarray}
    &&\mbox{\hspace{-4mm}}\tensor{R}{_\mu_\nu} \ \rightarrow \ \tensor{R}{_\mu_\nu}+\Omega^{-2}\left[2\left(\tensor{V}{_\mu_\nu}-\tensor{W}{_\mu_\nu}\right)-\tensor{g}{_\mu_\nu}\left(V+W\right)\right]\, ,\nonumber \\[2mm]
    &&\mbox{\hspace{-4mm}}\tensor{V}{_\mu_\nu}\ = \ \nabla_\mu\Omega\nabla_\nu\Omega\, ,~~~\tensor{W}{_\mu_\nu} \ = \ \Omega\nabla_\mu\nabla_\nu\Omega \, .
\end{eqnarray}
Here $V = \tensor{V}{_\mu^\mu}$ and $W = \tensor{W}{_\mu^\mu}$. The resulting expressions can be quite complicated in the general case, but become simpler as many of the spin coefficients vanish.  For example, $\Phi_{00}$, which is going to be important in the following considerations, transforms as
\begin{eqnarray}
&&\mbox{\hspace{-6mm}}\Phi_{00}\ \rightarrow \ \Omega^{-2}\Phi_{00} \ + \ \Omega^{-4}\left[2\mathrm{D}\Omega\mathrm{D}\Omega \ - \ \Omega\mathrm{D}\mathrm{D}\Omega \right. \nonumber 
\\[2mm]
&&\mbox{\hspace{5mm}}\left.+ \ \Omega\left(\epsilon \ + \ \bar{\epsilon}\right)\mathrm{D}\Omega\ 
 - \ \Omega\bar{\kappa}\delta\Omega\ - \ \Omega\kappa\bar{\delta}\Omega\right].
\end{eqnarray}
This can be simplified by choosing $l_\mu$ such that $\kappa=\epsilon+\bar{\epsilon}=0$. 
We see that by an appropriate choice of $\Omega$, $\Phi_{00}$ can be greatly simplified and often even nullified

{\em Fourth}, the spin coefficients transform as 
\begin{eqnarray}
    &&\kappa \ \rightarrow \ \Omega^{-1}\kappa\, ,\;\;\;\; \tau \ \rightarrow \ \Omega^{-1}\tau \ - \ \Omega^{-2}\delta\Omega\, , \\[2mm]
    &&\sigma \ \rightarrow \ \Omega^{-1}\sigma\, , \;\;\;\; \rho \ \rightarrow \ \Omega^{-1}\rho \ - \ \Omega^{-2}\mathrm{D}\Omega\, .~~~~~~
\end{eqnarray}
The same rule as for $\kappa, \tau, \sigma$, and $\rho$ applies to $\nu$, $\pi$, $\lambda$, and $\mu$, respectively. The third set transforms as
\begin{eqnarray}
    &&\epsilon \ \rightarrow \ \Omega^{-1}\epsilon \ + \ \frac{1}{2}\Omega^{-2}\mathrm{D}\Omega\, ,\\[2mm]
    &&\gamma \ \rightarrow \ \Omega^{-1}\gamma \ + \ \frac{1}{2}\Omega^{-2}\Delta\Omega\, ,\\[2mm]
    &&\beta \ \rightarrow \ \Omega^{-1}\beta\ - \ \frac{1}{2}\Omega^{-2}\delta\Omega\, ,\\[2mm]
    &&\alpha \ \rightarrow \ \Omega^{-1}\alpha \ - \ \frac{1}{2}\Omega^{-2}\bar{\delta}\Omega\, .
\end{eqnarray}
Importantly, $\kappa$ and $\sigma$ (and so also $\nu$ and $\lambda$) scale trivially. This agrees well with their physical interpretation, since $\kappa$ is zero if and only if $l_\mu$ is tangent to a null geodesic, and null geodesics are preserved by conformal transformations. Similarly, the quantity $\epsilon+\bar{\epsilon}$ measures how much the parameterization of the geodesic differs from the affine one.  We see that there is always a conformal transformation that brings us to an affinely parametrized frame.  As a consequence, the Goldberg--Sachs theorem also holds for spacetimes conformal to Ricci flat ones, which comprise a large family of solutions of WCG.

The Bach equations in the NP formalism were found in~\cite{Sv} and are summarized in Appendix~\ref{eq:NPBach}. Especially in the case of type II solutions (of which type D is a subset), a significant simplification will occur. The only nonzero NP-Weyl scalars are going to be $\Psi_2$, $\Psi_3$ and $\Psi_4$. Further significant simplification occurs in the case $\kappa=0$. Even though we do not know whether the Goldberg--Sachs theorem holds in general in WCG, $\kappa$ can often be set to zero by a suitable choice of the NP tetrad based on knowledge of null geodesic congruences, which can be guessed from the symmetry of the spacetime being considered (e.g. $l_\mu$ pointing in the radial direction in a spherically symmetric spacetime). 
The NP-Ricci scalar $\Phi_{00}$ is a gauge dependent quantity that can often be eliminated by a conformal transformation without affecting $\kappa$. $\epsilon$ can be set to zero by a suitable tetrad rotation which does not affect $l_\mu$. Finally, we can use one of the NP equations coming from Ricci identities, namely
\begin{eqnarray}
    \label{np:Ricci1}
    \mathrm{D}\rho \ - \ \bar{\delta}\kappa&=&\left(\rho^2 \ + \
    \sigma\bar{\sigma}\right) \ + \ \left(\epsilon\ + \ \bar{\epsilon}\right)\rho \ - \ \bar{\kappa}\tau\nonumber \\[2mm]&-&\kappa\left(3\alpha \ + \ \bar{\beta} \ - \ \pi\right)\ + \ \Phi_{00}\, ,
\end{eqnarray}
to trade $\sigma\bar{\sigma}$ for an expression containing only $\rho$. The null geodesics can be used to (locally) define a new coordinate $x$ such that $D\phi=\partial_x\phi$ for any scalar $\phi$. As a consequence, in Petrov type II spacetimes, one of the Bach equations reduces to 
\begin{eqnarray}
\frac{\partial^2 \Psi_2}{\partial x^2} &-& 6\frac{\partial}{\partial x}\left(\rho\Psi_2\right) \ + \ 12\rho^2\Psi_2 \ + \  {\rm c.c.}  \ = \  0\, , 
\end{eqnarray}
under a suitable choice of the conformal factor. This expression depends only on $\rho$ and if $\rho$ takes a simple form, as we will show in the next section, the resulting second order equation is easily solvable. Alternatively we could use \eqref{np:Ricci1} to express $\Phi_{00}$ using $\rho$ and $\sigma$ without fixing a gauge, a step that can be useful when $\sigma=0$.
}

{\cdblue
%
{\em Finally}, we may also mention the phenomenologically important case of Petrov type N solutions (which are not discussed in this paper), corresponding to spacetimes where there exists a null vector $k^\mu$ such that
\begin{eqnarray}
    \tensor{C}{_\mu_\nu_\alpha_\beta}k^{\mu} \ = \ 0\, .
\end{eqnarray}
If this vector is chosen to be one of the real NP tetrad vectors, let us say $l$, then $B_{ll}=B_{ln}=B_{lm}=0$. Which, again substantially simplifies the analysis. Petrov ``type-N''  solutions in WCG will be discussed in our subsequent paper. }

\section{Some exact solutions}
\label{Sec.3}

In this section we will derive and discuss the most important class of (electro) vacuum Petrov ``type-D'' solutions in WCG, namely static, spherically symmetric metrics. Apart from the well known black hole solutions, a new exotic wormhole solution will be presented.

\subsection{General considerations}
\label{subb section II A}

We begin with a metric for a static spherically symmetric source. In this case the line element has the form
\begin{equation}
    ds^2 \  = \ -\tilde{A}(\rho)dt^2 \ + \ \tilde{B}(\rho)d\rho^2 \ + \ \rho^2d\Omega^2\, .
    \label{III.A.13.cc}
\end{equation}
Due to the conformal invariance of the Bach equation, it is redundant to have two independent functions $\tilde{A}(\rho)$ and $\tilde{B}(\rho)$. Only one is truly independent, the other is determined by the choice of a conformal gauge. To see this, let us rewrite~(\ref{III.A.13.cc}) as
\begin{equation}
    ds^2 \ = \ \frac{\rho^2}{L(r)^2}\left[-A(r)dt^2  +  B(r)\ \! dr^2 +  L^2(r) \ \!  d\Omega^2\right] ,
\end{equation}
where
\begin{eqnarray}
    \rho \ = \ \rho(r)\, ,\;\;\;\; A\ = \  \frac{L^2 \tilde{A}}{\rho^2}\, ,\;\;\;\; B \ = \ \frac{L^2 \tilde{B}\left(\rho^{\prime}\right)^2}{\rho^2}\, .~~~
\end{eqnarray}
Here $L(r)$ is an arbitrary differentiable function from $\mathcal{C}^2$. 
A choice of $L(r)$ other than $L(r)=r$ will allow us to address, for example, exotic wormhole spacetimes (see section~\ref{Sec.III.C.cv}).
Note that each choice of $\rho(r)$ as a function of the new radial coordinate $r$ will constrain the new functions $\tilde{A}$ and $\tilde{B}$. 
The selection of $\rho(r)$ and $L(r)$ represents the choice of a gauge, that is, a selection of the function $\Omega$. 
One way to describe the gauge choice is to write one of the functions, say $A(r)$, as
\begin{equation}
    A \ = \ \tilde{\xi}(r) \ = \ \tilde{\xi}(r)\frac{B}{B}\, ,
\end{equation}
to get the relation
\begin{equation}
    AB \ = \ \xi(r) \ = \ \tilde{\xi}(r)B\, ,
\end{equation}
where the choice of $\xi(r)$ represents our gauge choice. The relation between $\xi$ and $\rho$ (for fixed $L$) is then
\begin{eqnarray}
    \frac{1}{\rho} \ = \ -\int\frac{\sqrt{\xi} \ \! dr}{L^2\sqrt{\tilde{A}\tilde{B}}}\, .
    \label{24nm}
\end{eqnarray}
Thus, given $\tilde{A}$ and $\tilde{B}$, we can choose functions $\xi(r)$ and $L(r)$ (that is, a gauge transformation) such that we can pass to an equivalent metric-field configuration in which $A(r)$ and $B(r)$ are manifestly dependent functions.

To find a suitable gauge for our calculations, we consider the following NP null tetrad 
\begin{eqnarray}
    &&l \ = \ -\sqrt{\frac{B}{A}}\partial_t \ - \
    \partial_r\, ,\;\;\;\; n \ = \ -\frac{\partial_t}{2\sqrt{AB}} \ + \ \frac{\partial_r}{2B}\, , \nonumber \\[2mm]
    &&m \ = \ -\frac{\partial_\theta}{\sqrt{2}L} \ - \ i\frac{\partial_\varphi}{\sqrt{2}L\sin(\theta)}\, .
    \label{III.!9.cc}
\end{eqnarray}\\[-1mm]
This choice of tetrad is motivated by the desire to reduce one of the directional derivative operators to $\mathrm{D}=-\partial_r$. We will see that in such a case the Bach equations in spherical coordinates are easier to handle.
It can be checked that the only non-zero spin coefficients for the above choice of tetrad are 
\begin{eqnarray}
    &&\alpha \ = \ -\beta \ = \ \frac{\sqrt{2}}{4L \tan\left(\theta\right)}\, ,\nonumber \\[2mm]
    &&\gamma \ = \ \frac{B^{\prime}}{8B^2} \ - \ \frac{A^{\prime}}{8AB}\, ,\nonumber \\[2mm]
    &&\epsilon \ = \ -\frac{B^{\prime}}{4B} \ - \ \frac{A^{\prime}}{4A}\, ,\nonumber \\[2mm]
    &&\rho \ = \  \frac{L^{\prime }}{L}\, ,\;\;\;\;
    \mu \ = \  \frac{BL^{\prime}}{2L}\, ,
    \label{IV.26.cv}
\end{eqnarray}
and the only non-zero curvature scalars are $\Lambda$, $\Phi_{00}$, $\Phi_{11}$, $\Phi_{22}$ and $\Psi_2$. This shows that all spherically symmetric solutions are of Petrov ``type-D'' character~\cite{Penrose:60,Chandra}, see also Appendix~\ref{Appendix A}.
\begin{widetext}
The Bach vacuum equations reduce to (cf. Appendix~\ref{Appendix A})\\
\begin{eqnarray}
    \label{eq:Bach1}
  &&\mathrm{D}\mathrm{D}\Psi_2 \ - \ 6\rho \mathrm{D}\Psi_2 \ + \ 9\rho^2\Psi_2 \ - \ 3\Psi_2 \mathrm{D}\rho \ + \ \Phi_{00}\Psi_2\ = \ 0\, ,   \\[2mm]
    \label{eq:Bach2}
        &&\mathrm{D}\Delta\Psi_2 \ + \ 2\mu \mathrm{D}\Psi_2 \ - \ 2\rho\Delta\Psi_2 \ - \ 3\rho\mu\Psi_2 \ + \ 3\Psi_2 \mathrm{D}\mu  \ + \ 2\Phi_{11}\Psi_2 \ = \ 0\, .
\end{eqnarray}
\end{widetext}
From the choice (\ref{III.!9.cc}) we might note that the corresponding $B_{nn}$ component of the 
Bach equation need not be considered, since $B_{nn}$ is
is trivially proportional to $B_{ll}$.

Most importantly, \eqref{eq:Bach1} is very simple because there is no $A$ or $B$ dependence in $\rho$ and since we consider a static solution, $\mathrm{D}$ reduces to $-\partial_r$. The Ricci NP scalar appearing in the equations is equal to
\begin{eqnarray}
    \label{eq:Phi00}
    \Phi_{00}&=&-\frac{L^{\prime\prime}}{L} \ + \ \frac{L^{\prime}}{2L}\frac{AB^{\prime} \ + \ BA^{\prime}}{AB}\nonumber \\[2mm]
    &=& - \ \frac{L^{\prime\prime}}{L} \ + \ \frac{L^{\prime}}{2L} \ \! (\log \xi)^{\prime}\, .
\end{eqnarray}
A convenient gauge choice would be the one in which~(\ref{eq:Phi00}) takes a simple form. It will be seen shortly that by choosing  a suitable $\xi$, we can reduce the original problem of solving a fourth-order equation to a series of two second-order equations.
Note that Eq.~\eqref{eq:Bach1} does not contain any term explicitly dependent on $B$ or $A$, and therefore $\Psi_2$ could be solved by methods and techniques used in the theory of second-order linear differential equations. 

For example, 
from~\eqref{eq:Phi00} we see that the simplest gauge for a fixed $L=r$ is $\xi=\rm{const}$. Since the multiplicative constant can always be absorbed by a rescaling of the time coordinate, we choose the gauge $\xi = 1$, i.e. $A=B^{-1}$. 
In this case, the NP null tetrad~(\ref{III.!9.cc}) reduces to
\begin{eqnarray}
    &&l \ = \ -\frac{\partial_t}{A(r)} \ - \
    \partial_r\, ,\;\;\;\; n \ = \ -\frac{\partial_t}{2} \ + \ \frac{A(r)\partial_r}{2}\, , \nonumber \\[2mm]
    &&m \ = \ -\frac{\partial_\theta}{\sqrt{2}L(r)} \ - \ i\frac{\partial_\varphi}{\sqrt{2}L(r)\sin(\theta)}\, .
\end{eqnarray}
The ensuing non-zero curvature scalars acquire the form
\begin{widetext}
\begin{eqnarray}
    &&\mbox{\hspace{-8mm}}\Phi_{00} \ = \ -\frac{L^{\prime\prime}}{L}\, , \;\;\;\;
    \Phi_{11} \ = \  \frac{L^2 A^{\prime\prime} \ - \ 2\left(L^{\prime}\right)^2A \ + \ 2}{8L^2}\, , \;\;\;
    \Phi_{22}\ = \ -\frac{A^2 L^{\prime\prime}}{4L}\, , \nonumber \\[2mm]
&&\mbox{\hspace{-8mm}}\Lambda \ = \ \frac{-4ALL^{\prime\prime}-2A\left(L^{\prime}\right)^2-L^2A^{\prime\prime}-4LA^{\prime}L^{\prime}+2}{24L^2}\, , \;\;\;\; \Psi_2 \ = \
\frac{-2ALL^{\prime\prime}+2A\left(L^{\prime}\right)^2+L^2A^{\prime\prime}-2LL^{\prime}A^{\prime}-2}{12L^2}\, .
\label{III.25.hj}
\end{eqnarray}
\end{widetext}
As an aside, we note that the simplest gauge to solve for a black hole ($L=r$) is the gauge choice $\xi = 1$, which corresponds to $\Phi_{00}=0$. This is also the only gauge that contains the Schwarzschild solution, since Einstein's vacuum field equation solutions require $\Phi_{00}=0$.

\subsection{Mannheim--Kazanas black hole solution}\label{MK}

To obtain a solution for the WCG black hole in Schwarzschild coordinates, we fix $L(r)=r$. The Bach equation~\eqref{eq:Bach1} is no longer dependent on $B$ and reduces to the Cauchy--Euler differential aequation
\begin{equation}
    \label{eq:MKBlackHoleBachEquation1}
r^2\Psi_2^{\prime\prime} \ + \ 6r\Psi_2^{\prime} \ + \ 6\Psi_2 \ = \ 0\, ,
\end{equation}
which has the general solution~\cite{zvi}
\begin{equation}
    \label{eq:BlackHoleWeylScalar}
    \Psi_2 \ = \ \frac{c_1}{r^3} \ + \ \frac{c_2}{r^2}\, .
\end{equation}
To find $B$, we must first solve the equation determining $\Psi_2$ in terms of $B$, see Eqs.~(\ref{III.25.hj}), and then check the remaining Bach equation (\ref{eq:Bach2}) for additional constraints on the integration constants. 

With the use of~\eqref{III.25.hj}, $B$ is given by the equation\footnote{Multiplicative factor of 12 can was absorbed into redefintion of $c_1$ and $c_2$.}
\begin{equation}
        r^2 A^{\prime\prime} \ - \ 2rA^{\prime} \ + \ 2A \ = \ \frac{c_1}{r} \ + \ c_2 \ + \ 2\, .
\end{equation}
This can be solved e.g. by the method of variation of parameters, where we can take $r$ and $r^2$ as independent solutions of the corresponding homogeneous equation (the Cauchy--Euler equation). The general solution is
\begin{equation}
    A \ = \ a \ + \ \frac{p}{r} \ + \ qr \ + \ \zeta r^2\, ,
    \label{III.29.cd}
\end{equation}    
where $a$, $p$, $q$ and $\zeta$ are integration constants. The algebraic constraint for these constants is given by the second non-trivial Bach equation~\eqref{eq:Bach2} and it reads
\begin{eqnarray}
3pq \ + \ 1 \ - \ a^2 \ = \ 0\, .
 \label{III.30.cd}
\end{eqnarray}
In particular, note that there is no constraint on $\zeta$.
As noted in Ref.~\cite{mannheim_b,mannheim_c,mannheim_a}, this is because in WCG, a given local gravitational source generates only first three terms in a gravitational potential~(\ref{III.29.cd}).
The term $\propto r^2$ corresponds to a trivial vacuum solution of WCG and hence does not couple to matter sources~\cite{mannheim_b,mannheim_c}. Solution~(\ref{III.29.cd}) together with~(\ref{III.30.cd}) 
includes as special cases Einstein spaces, namely the Schwarzschild solution
($q\!=\!\zeta\! = \!0$) and the Schwarzschild--de-Sitter solution ($q\!=\!0$),
where the latter does not require the presence of a cosmological
constant as Einstein's gravity does. Furthermore, it can be shown that the linear term in~(\ref{III.29.cd}) can be gauged away by a suitable Weyl rescaling~\cite{Schmidt}. 

%

 It can be checked that (\ref{III.29.cd}) together with (\ref{III.30.cd}) coincide with the Mannheim--Kazanas solution~\cite{MK2}.

\subsection{Charged black hole solution} \label{Sec.III.C.cv}

The solution can easily be extended to include also the electromagnetic field. In particular, we will focus on a black hole in WCG with a point-like static charge $Q$. No spin and magnetic dipole moment will be considered. The electromagnetic four-potential is therefore a Coulomb potential. This will lead to a WCG analog of
the Reissner--Nordstr\"{o}m solution.  

In the NP formalism the electromagnetic field is characterized by three complex Maxwell scalars~\cite{Taukolsky}
\begin{eqnarray}
\phi_0 &=& F_{\mu\nu}l^\mu m^\nu\, ,\;\;\; \phi_1 \ = \ \frac{1}{2} F_{\mu\nu} \left(l^\mu n^\nu \ + \ \bar{m}^\mu m^\nu  \right) \, , \nonumber \\[2mm]
\phi_2 &=& F_{\mu\nu} \bar{m}^\mu n^\nu\, ,
\end{eqnarray}
or, equivalently
\begin{eqnarray}
F_{\mu\nu} &=& 2 \left[\phi_1(n_{[\mu}l_{\nu]} \ + \ m_{[\mu}\bar{m}_{\nu]})\right. \nonumber \\[2mm] &&\left.+ \ \phi_2 l_{[\mu}m_{\nu]} \ + \  \phi_{0}\bar{m}_{[\mu}n_{\nu]} \right] \ + \ \mbox{c.c.}\, .
\label{B.38.cv}
\end{eqnarray}
By analogy with Einstein spaces, we might again expect the final spacetime to be of a Petrov ``type-D'', in which case the Maxwell equations can be written in NP formalism in the form~\cite{Taukolsky}  
\begin{eqnarray}
        \label{eq:MaxwellNP1}
&&(D\ - \ 2 \rho) \phi_1 \ - \ \left(\bar{\delta} \ + \  \pi \ - \ 2 \alpha\right) \phi_0 \ = \ 2 \pi J_l\, , \\[2mm]
&&(\delta \ - \ 2 \tau) \phi_1 \ - \ (\Delta \ + \ \mu \ - \  2 \gamma) \phi_0  \ = \ 2 \pi J_m\, , \\[2mm]
&&(D \ - \ \rho \ + \ 2 \epsilon) \phi_2 \ - \ \left(\bar{\delta} \ + \ 2 \pi\right) \phi_1 \  = \ 2 \pi J_{\bar{m}}\, , \\[2mm]
&&\label{eq:MaxwellNP2}
(\delta \ - \ \tau \ + \ 2 \beta) \phi_2 \ - \ (\Delta \ + \ 2 \mu) \phi_1 \  = \ 2 \pi J_n\, ,~~~~~~~~~
\end{eqnarray}
where $J$ is the vector current represented in the local basis $\{l^{\mu}, n^{\mu}, m^{\mu}, \bar{m}^{\mu}\}$.
For the situation at hand, i.e. exterior solutions, we consider only the case where there are no currents. For a static spherically symmetric spacetime the Maxwell scalars are allowed to be dependent only on $r$. Under this assumption, Eq.~\eqref{eq:MaxwellNP1} forces $\phi_0=0$, because $\alpha$ is  $\theta$ dependent, cf.~(\ref{IV.26.cv}). The same argument applies to \eqref{eq:MaxwellNP2} and $\phi_2$. As a result, Eqs.~(\ref{eq:MaxwellNP1})-(\ref{eq:MaxwellNP2}) reduce to a single equation
\begin{eqnarray}
    \phi_1^{\prime} \ + \ \frac{2}{r}\phi_1 \ = \ 0\, .
\end{eqnarray}
A simple separation of variables yields the general solution $\phi_1=C/r^2$, where $C$ is an integration constant which has to be determined through other physical assumptions. Note in particular that, since only $\phi_1$ survives, it is due to~(\ref{B.38.cv}) that $\phi_1$ is built up solely from the physically measurable $\tensor{F}{_\mu_\nu}$ and thus the integration constant $C$ is not related to a gauge choice for the electromagnetic four-potential. In order to determine $C$, we compute the EM field tensor, which from~(\ref{B.38.cv}) reads
\begin{eqnarray}
    \label{eq:EMFieldTensor}
    \tensor{F}{_\mu_\nu}=\begin{pmatrix}
        0 & -\frac{2C}{r^2} & 0 & 0 \\
        \frac{2C}{r^2} & 0 & 0 & 0 \\
        0 & 0 & 0 & 0 \\
        0 & 0 & 0 & 0 \\
    \end{pmatrix}.
\end{eqnarray}
Since the Maxwell equations used are independent of $B$,~\eqref{eq:EMFieldTensor} must also apply to flat spacetime. Therefore in order to restore the electric field of a point particle we should set $C=Q/2$. The four-potential is then 
\begin{equation}
    \label{eq:MKMaxwellAnsatz}
    \mathcal{A} \ = \ \frac{Q}{r}dt\ + \ d\omega \, ,
\end{equation}
where $\omega$ is an arbitrary function corresponding to the gauge chosen. As in (\ref{eq:Bach1})-(\ref{eq:Bach2}), only the $B_{ll}$ and $B_{ln}$ components of the Bach equation are relevant for spherically symmetric solutions of the Petrov ``type D''. Together with the resulting energy-momentum tensor components  
\begin{eqnarray}
    T_{ll} \ = \ 0\, ,\;\;\;\;\;T_{ln}\ = \ \frac{Q^2}{2r^4}\, ,
\end{eqnarray}
we obtain the inhomogeneous equations of the motion. Note that the ``$ll$''-component equation yields exactly the  Mannheim--Kazanas solution
\begin{equation}
   A \ = \ a \ + \ \frac{p}{r} \ + \ qr \ + \ \zeta r^2\, ,
   \label{III.47.cc}
\end{equation}
as we had in (\ref{III.29.cd}). The second --- ``$ln$''-component equation of motion provides a modified to algebraic constraint [cf. Eq.~(\ref{III.30.cd}) for the Schwarzschild-like case]
\begin{eqnarray}
 3pq+1 \ - \ a^2 \ = \ \frac{3G_{\rm{w}}^2Q^2}{2}\, .
\label{C.22.cc}
\end{eqnarray}
By setting $c_1=c_2=0$ in \eqref{eq:BlackHoleWeylScalar}, the most general conformally flat solution is given by
\begin{equation}
    A\ = \ 1 \ + \ a_1r \ + \ a_2r^2\, .
\end{equation}
where $a_1$ and $a_2$ are arbitrary integration constants. Together with (\ref{C.22.cc}) this implies that the most general conformally flat solution must correspond to $Q=0$, or in other words, Reissner--Nordstr\"{o}m solution cannot be conformally flat.

\section{General solution \label{Sec.V.cd}}

Exact solutions from Section~\ref{Sec.3} can be obtained from a more general unifying framework. In this section we will discuss such a framework and, in addition, derive yet another class of exact solutions --- wormhole solutions.

We start by rewriting, for arbitrary $\xi$, the non-zero spin coefficients (\ref{IV.26.cv}) as
\begin{eqnarray}
    &&\alpha \ = \  -\beta \ = \ \frac{\sqrt{2}}{4 L \tan{\left(\theta \right)}}\, ,\nonumber\\[2mm]
    &&\gamma \ = \  \frac{A \xi^{\prime} - 2 \xi A^{\prime}}{8 \xi^{2}}\, , \;\;\;\; \epsilon \ = \  - \frac{\xi^{\prime}}{4 \xi}\, , \nonumber \\[2mm]
    &&\rho \ = \  \frac{ L^{\prime}}{L}\, ,\;\;\;\; \mu \ = \  \frac{AL^{\prime}}{2 L \xi}\, .
\end{eqnarray}
With these the Bach tensor component $B_{ll}$ acquires the form
\begin{eqnarray}
    B_{ll} &=&\ 2\left[\Psi_2^{\prime\prime} \ - \ \frac{\Psi_2^{\prime} \xi^{\prime}}{2 \xi} \ + \  \frac{2 \Psi_2 L^{\prime\prime}}{L} \ - \  \frac{\Psi_2 L^{\prime} \xi^{\prime}}{L \xi} \right. \nonumber \\[2mm]
    &&+ \ \left.\frac{6 L^{\prime} \Psi_2^{\prime}}{L} \ + \  \frac{6 \Psi_2 \left(L^{\prime}\right)^{2}}{L^{2}}\right].
\end{eqnarray}
By writing $\Psi_2(r)$ in the form $\Psi_2(r)=\omega(r)/L^2(r)$, $B_{ll}$ can be rewritten as
\begin{eqnarray}
    B_{ll} \ = \ 2\left[\frac{\omega^{\prime\prime}}{L^{2}} \ - \ \frac{\omega^{\prime} \xi^{\prime}}{2 L^{2} \xi} \ + \ \frac{2 L^{\prime} \omega^{\prime}}{L^{3}}\right].
\end{eqnarray}
Since this expression contains only the derivatives of $\omega$, the resulting expression is a separable first-order ordinary differential equation (ODE) for $\omega^{\prime}$, and the full solution for $\Psi_2$ of the vacuum equation $B_{ll} = 0$ can be thus given by
\begin{eqnarray}
    \Psi_2 \ = \  \frac{C_{1}}{L^2} \ + \ \frac{C_{2}}{L^2} \int \frac{\sqrt{\xi}}{L^{2}}\, dr\, .
\end{eqnarray}
The next step is to express non-zero curvature scalars $\Psi_2$ from (\ref{III.16.cc}) in terms of $A$, $L$ and $\xi$. A simple algebra reveals that the resulting expression is
\begin{eqnarray}
    \Psi_2 \!&=& \!\frac{- \ 4 \left(A L^{\prime\prime} \  +  \ A^{\prime} L^{\prime}\right) L \xi \ + \ 2 A L L^{\prime} \xi^{\prime} \ + \ 4 A \xi \left(L^{\prime}\right)^{2}  }{24 L^{2} \xi^{2}}\nonumber \\[2mm] &&+ \ \frac{2 L^{2} \xi A^{\prime\prime} \ - \  L^{2} A^{\prime} \xi^{\prime} \ - \  4 \xi^{2}}{24 L^{2} \xi^{2}}\,.~~~~~~~
\end{eqnarray}
The substitution $A=\Theta L^2$ brings it into a more manageable form
\begin{eqnarray}
    \Psi_2 \ = \ \frac{2 \left(L \Theta^{\prime\prime} \ + \  2 L^{\prime} \Theta^{\prime}\right) L^{3} \xi \ - \  L^{4} \zeta^{\prime} \xi^{\prime} \ - \  4 \xi^{2}}{24 L^{2} \xi^{2}}\, , ~~~
\end{eqnarray}
which again contains only the derivatives of $\Theta$ and leads thus to a first order ODE for $\Theta^{\prime}$. Because of linearity the solution to the full nonhomogeneous case can be found by variation of parameters. The homogeneous solution is
\begin{eqnarray}
    A \  = \ C_3L^2 \ + \ C_4L^2\int \frac{\sqrt{\xi}}{L^2} dr\, .
\end{eqnarray}
By setting $C_1=C_2=0$ we can find the most general conformally flat solution to be
\begin{widetext}
\begin{eqnarray}
   A \ = \ C_3L^2 \ + \ C_4L^2\int \frac{\sqrt{\xi}}{L^2} dr \ + \ 2L^2\int\left(\int\frac{\sqrt{\xi}}{L^2}dr\right)\frac{\sqrt{\xi}}{L^2}dr\, .
\end{eqnarray}
Using the method of variation of parameters, the particular solution of the non-homogeneous equation for $\Psi_2$ is obtained in the standard way from the previous result. Consequently, the most general static spherically symmetric solution of the Bach equations in arbitrary gauge is of the form
\begin{eqnarray}
    A &=& \left[C_3 \ + \  \int \frac{\left(C_4 \ + \  2 \int \frac{\sqrt{\xi}}{L^{2}}\, dr\right) \sqrt{\xi}}{L^{2}}\, dr\right] L^{2} \ + \  12 \left[\left(\int \frac{\sqrt{\xi}}{L^{2}}\, dr\right) \int \frac{\left(C_1 \ + \  C_2 \int \frac{\sqrt{\xi}}{L^{2}}\, dr\right) \sqrt{\xi}}{L^{2}}\, dr \right.
    \nonumber \\[2mm]&&- \ \left. \int \frac{\left(C_1 \ + \  C_2 \int \frac{\sqrt{\xi}}{L^{2}}\, dr\right) \sqrt{\xi} \int \frac{\sqrt{\xi}}{L^{2}}\, dr}{L^{2}}\, dr\right] L^{2}\,.
    \label{58.bb}
\end{eqnarray}
If we use a simple integral identity from classical calculus
\begin{eqnarray}
\int_{a_0}^r dr_1f(r_1)\int_{a_0}^{r_1} dr_2 f(r_2) \ldots  \int_{a_0}^{r_{n-1}} dr_n f(r_n)  \ = \ \frac{1}{n!} \left(\int_{a_0}^{r} dr' f(r') \right)^n\!,~~
\label{59.kj}
\end{eqnarray}
(which is valid for any constant $a_0$) we can rewrite (\ref{58.bb}) formally as
\begin{eqnarray}
    \label{eq:GeneralSolutionNice}
    A \ = \  L^2\left(a I^2 -\   \ p I^3 \ - \ q I \ + \ \zeta\right)\, ,\;\;\;\;
    I \ = \  \int \frac{\sqrt{\xi}}{L^2} dr\, ,
    \label{59.nm}
\end{eqnarray}
where  the integration constants are defined so as to match the Mannheim--Kazanas solution~(\ref{III.29.cd}) in the  Mannheim--Kazanas gauge $L= r$ and $\xi =1$. 

Finally, Eq.~\eqref{eq:Bach2} provides a constraint for the integration constants $C_1, C_2, C_3$ and $C_4$, namely
\begin{eqnarray}
    \frac{C_4 C_2}{2}\ - \  3 C_1^{2} \ - \  C_1  \ - \  3 C_2^{2} \left(\int \frac{\sqrt{\xi}}{L^{2}}\, dr\right)^{2} \ + \  6 C_2^{2} \int \frac{\sqrt{\xi} \int \frac{\sqrt{\xi}}{L^{2}}\, dr}{L^{2}}\, dr \ = \ 0\, .
    \label{81}
\end{eqnarray}
\end{widetext}
Notice that the last two terms in (60) cancel each other out due to the identity~(\ref{59.kj}) leaving behind only the condition
\begin{eqnarray}
    \frac{C_4 C_2}{2}\ - \  3 C_1^{2} \ - \  C_1  \ \ = \ 0\, .
    \label{81b}
\end{eqnarray}
Constraint~(\ref{81b}) is analogous to the constraint~(\ref{III.30.cd}) to which it reduces when $L=r$ and $\xi =1$, provided we identify
\begin{eqnarray}
&&a \ = \ 6C_1 \ + \ 1\, , \;\;\;\; p \ = \ -2C_2\, ,\nonumber \\[2mm] 
&&q \ = \ -C_4\, ,\;\;\;\; \zeta \ = \ C_3\, .  
\label{63.kk}
\end{eqnarray}
Here, the integration constant in $I$ was implicitly assumed to be zero.
Note that for~(\ref{63.kk}) we have that~(\ref{58.bb}) correctly reduces to~(\ref{III.47.cc}). In addition, similarly to the constant $\zeta$ in (\ref{III.29.cd}) and (\ref{III.47.cc}), $C_3$ is not restricted by the equation for $B_{ln}=0$. This is because $L^2$ alone is a solution of the vacuum equation $B_{ln} = 0$, and therefore the term $L^2$ is always allowed in the solution, even if we would be dealing with a non-vacuum situation (e.g. Reissner--Nordstr\"{o}m type solution). 
%
%
%
By choosing the gauge $\xi=\left(L^{\prime}\right)^2$ with $L(0) =0$, we see that we get back the Mannheim--Kazanas solution with a reparametrized radial coordinate $L(r)$ instead of $r$.

{\cdblue

Let us add a final remark concerning the integration constants in~\eqref{eq:GeneralSolutionNice}. Petrov type II spacetimes, of which type D is a subclass, are equivalently characterized by the existence of a null vector $k^{\mu}$ such that~\cite{Chandra}
\begin{eqnarray}
    \tensor{C}{_\mu_\nu_\alpha_\beta}k^{\mu}k^{\alpha} &=& \alpha k^{\mu}k^{\alpha}\, , \nonumber \\[2mm]
    ^{*}\tensor{C}{_\mu_\nu_\alpha_\beta}k^{\mu}k^{\alpha} &=& \beta k^{\mu}k^{\alpha}\, ,\;\;\;\; \alpha\beta \ \neq \ 0\, .
\end{eqnarray}
For a type D spacetime there exist two linearly independent vectors $k^{\mu}$ and $k^{\prime\mu}$ which satisfy this condition with generally different constants. These vectors can be used to form a tetrad (this choice was not pursued in our calculations), and the respective projections of the Bach tensor then become
\begin{eqnarray}
    &&B_{kk} \ = \ \alpha\left[\nabla^\mu\nabla^\nu\left(k_\mu k_\nu\right) \ - \ \frac{1}{2}\tensor{
R}{^\mu^\nu}k_\mu k_\nu\right]\, , \label{eq:TypeD1} \\[2mm]
    &&B_{k^{\prime}k^{\prime}} \ = \ \beta\left[\nabla^\mu\nabla^\nu\left(k^{\prime}_\mu k^{\prime}_\nu\right) \ - \ \frac{1}{2}\tensor{
R}{^\mu^\nu}k^{\prime}_\mu k^{\prime}_\nu\right]\, .~~~~~
\label{eq:TypeD2}
\end{eqnarray}\\
Unlike the general form of the Bach equations, \eqref{eq:TypeD1} and \eqref{eq:TypeD2} are of second order only. The unknown functions appearing in these equations can only have two integration constants. The other constants appear through the choice of the conformal factor and can be gauged away.

}

\subsection{Wormhole solution \label{wormhole-a}}

The general solution~(\ref{58.bb}) allows to discuss wormhole solutions in WCG.
To illustrate this, let us concentrate on a class of  spherically symmetric wormhole solutions described by the line element 
\begin{equation}
    ds^2\ = \ -A(r)dt^2 \ + \ B(r)dr^2 \ + \ L(r)^2d\Omega^2\, ,
    \label{V.68.kl}
\end{equation}
where the $r$ coordinate is analytically extended to negative values, and  $L(0)=L_{\rm{min}}\neq 0$.

A simple class of wormhole solutions can be obtained by considering $\xi=1$ and $L=\sqrt{r^2+r_0^2}$ with $r_0 >0$. In this case we have
\begin{widetext}
\begin{eqnarray}
    A &=& \left(r^2+r_0^2\right) \left[{a} \left(\int_{a_0}^r \frac{\sqrt{\xi}}{L^2} dr' \right)^2 \ - \ {p}\left(\int_{a_0}^r \frac{\sqrt{\xi}}{L^2} dr' \right)^3  \ - \ {q} \int_{a_0}^r \frac{\sqrt{\xi}}{L^2} dr'   \ + \ \zeta\right] \nonumber \\[2mm] 
    &=& \left(r^2+r_0^2\right)\left\{{a}\left[\frac{\arctan\left(\frac{r}{r_0}\right)}{{r_0}}-\frac{\pi}{2r_0}\right]^2 \ - \ {p}\left[\frac{\arctan\left(\frac{r}{r_0}\right)}{{r_0}}-\frac{\pi}{2r_0}\right]^3 \ - \ {q}\left[\frac{\arctan\left(\frac{r}{r_0}\right)}{{r_0}}-\frac{\pi}{2r_0}\right] \ + \ \zeta\right\} .~~~~~
\end{eqnarray}
\end{widetext}
Here we have set $a_0 = \infty$ to ensure that the resulting formula for $r\gg r_0$ reduces to the Mannheim--Kazanas formula~(\ref{III.29.cd}). After selecting 
\begin{eqnarray}
a \ = \ - 1\, ,\;\;\;\; q \ = \ \frac{\pi}{r_0}\, ,\;\;\;\; p \ = \ 0\, ,  
\end{eqnarray}
we obtain that
\begin{eqnarray}
A \ = \ \left(r^2+r_0^2\right) \left[\frac{\pi^2 - 4 \arctan^2\left(\frac{r}{r_0}\right)}{4r_0^2} \ + \ \zeta  \right] .
\label{66.dd}
\end{eqnarray}
This solution is parity invariant
(as expected for a 
two-way traversable wormhole) and in the $|r| \rightarrow \infty$ limit it approaches the Mannheim--Kazanas solution with the constraining condition~(\ref{III.30.cd}).

Note also that for $\zeta > 0$ we have $A>0$, so that $B$ has no singularity, which can also be checked independently by computing the Kretschmann invariant. 
In turn this also implies that
the resulting wormholes have no event horizon.
Again, this is to be expected for traversable wormholes, non-traversable wormholes (such as the famous Schwarzschild wormhole) have both a horizon and an anti-horizon (i.e. a past event horizon)~\cite{Thorne:88}.
It is interesting to observe that, contrary to Einstein's gravity, the WCG wormhole solution does not require ``exotic matter'' with negative energy to keep the wormhole throat open. In fact, the obtained two-parameter class of wormhole solutions (\ref{66.dd}) does not require any matter (exotic or non-exotic) for its stability, since it represents full-fledged  vacuum solutions of WCG.

We may observe that for $r_0 \ll |r|$ (i.e., wormhole throat vicinity) and $\zeta = (4-\pi^2)/(4r_0^2)$ we get $A = 1 + \mathcal{O}((r/r_0)^4)$ and so in the leading order behavior in $(r/r_0)^2$ our WCG wormhole solution coincides with the Ellis--Bronnikov wormhole known from
Einstein's gravity~\cite{Thorne:88,Ellis,Bronnikov}. In this context it is interesting to note that the Ricci NP scalar $\Phi_{00}$ for the above wormholes reads
\begin{eqnarray}
\Phi_{00}(r) \ = \ - \frac{r_0^2}{(r^2 + r_0^2)^2} \, ,
\end{eqnarray}
which is clearly different from zero. 
Since Einstein's vacuum field equation solutions require $\Phi_{00} = 0$, the wormhole solution~(\ref{66.dd}) does not exist in Einstein's gravity, and is thus a genuine non-trivial prediction of WCG.  

Strictly speaking, for a truly traversable wormhole, one would have to check that the solution is perturbatively stable against conformal perturbations (e.g. tidal effects). However, enforcing this requirement would inevitably involve time-dependent and non-spherical analysis, which is beyond the scope of this article.

In the presence of EM field a computation analogous to the Mannheim--Kazanas case reveals that only the algebraic condition has to be modified. The NP-Maxwell equations reduce to
\begin{eqnarray}
    \phi_1^{\prime} \ + \   
    \frac{2L^{\prime}}{L}\phi_1 \ = \ 0\, ,
\end{eqnarray}
and the solution is given by $\phi_1=C/L^2$. By setting $C=Q/2$ as before, we obtain
\begin{eqnarray}
    T_{ll} \ = \ 0\, ,\;\;\;\;\;T_{ln}\ = \ \frac{Q^2}{2L^4}\, .
\end{eqnarray}
The resulting algebraic constraint is 
\begin{eqnarray}
    1 \ + \ 3pq \ - \ a^2 \ = \ \frac{3G_{\rm{w}}^2Q^2}{2}\, .
\end{eqnarray}



\subsection{WCG and equivalence class of solutions}

The question naturally arises as to whether the wormhole solution from the previous section is simply not equivalent (under the Weyl transformation) to the Mannheim--Kazanas black hole solution. Since the physically observable propositions in WCG should be gauge independent, the latter would mean that the wormhole solutions obtained could not correspond to real traversable wormholes. However, it is easy to see from~(\ref{24nm})  
that there is no Weyl's $\Omega(r)$ connecting the Mannheim--Kazanas solution of Sec.~\ref{MK} with the wormhole solution of Sec.~\ref{wormhole-a}. Indeed, starting from the Mannheim--Kazanas solution, we have in~(\ref{III.A.13.cc}) that $\tilde{A}(\rho) = \tilde{B}^{-1}(\rho)$, on the other hand, the wormhole solution~(\ref{66.dd}) corresponds to $\xi = 1$ and $L = \sqrt{r^2 + r_0^2}$ with $r_0>0$. By~(\ref{24nm}), we may thus write
\begin{eqnarray}
 \frac{1}{\rho(r)} &=& -\int_{a_0}^r\frac{\sqrt{\xi} \ \! dr'}{L^2\sqrt{\tilde{A}\tilde{B}}} 
 \nonumber \\[2mm] 
 &=& 
 \frac{1}{r_0} \ \! {\arctan\left(\frac{a_0 \ - \ r}{r_0 \ + \  a_0r/r_0}\right) }\, .~~~~
    \label{24cc}
\end{eqnarray}
Obviously, this cannot be satisfied, since $\rho$ runs from $0$, while the right side is a bounded function for any $a_0$ and $r_0$. This is true even if we were working with the maximally extended version of the Mannheim--Kazanas spacetime (i.e. the version where $-\infty < \rho < \infty$).
Consequently, the two metric field configurations cannot be connected via any Weyl's $\Omega$ --- they are inequivalent in WCG.

{\cdblue
Alternatively, the above inequivalence can be deduced from the fact that the Weyl squared term is not only a scalar, but in four dimensions it is also invariant under Weyl transformations. So, in particular, if $C^2$ is singular in one spacetime, then there is no non-singular conformal factor $\Omega$ which would remove the singularity in another  Weyl-equivalent spacetime. In our case, for the solution~\eqref{eq:GeneralSolutionNice}, we have
\begin{eqnarray}
    &&\mbox{\hspace{-3mm}}\tensor{C}{_\mu_\nu_\alpha_\beta}\tensor{C}{^\mu^\nu^\alpha^\beta}\nonumber \\[2mm]
    &&\mbox{\hspace{-3mm}}= \ \frac{4 \left(a^{2} \ - \ 6 a p I \ - \  2 a \ + \ 9 p^{2} I^{2} \ + \ 6 p I \ + \  1\right)}{3 L^{4}}\, ,~~~~~
\end{eqnarray}
where the same notation and logic regarding the integration constants as in \eqref{eq:GeneralSolutionNice} applies. This expression is singular only at points where $L=0$ (which includes  the Mannheim--Kazanas spacetime). Solutions with $L>0$ everywhere, such as the WCG wormhole solution~(\ref{66.dd}), cannot therefore be Weyl equivalent to the Mannheim--Kazanas solution. 
}

{\cdblue 
Note that the gauge inequivalence of the Mannheim--Kazanas black hole solution and the WCG wormhole solution can also be justified by topological (rather than geometric) arguments. To illustrate this, we will utilise two topological methods: {\em homotopy theory} and the {\em Betti numbers}.
First, we compare the homotopic properties of the two types of manifolds.
To this end, we start with the Mannheim--Kazanas case. 
The Mannheim--Kazanas black hole has a similar topology to the Schwarzschild black hole, but the global structure is modified by the linear and quadratic terms.  In addition, the Mannheim--Kazanas solution can have multiple horizons, depending on the values of the parameters $a$, $p$, $q$, and $\zeta$. All such prospective horizons are only coordinate singularities, because the Mannheim-Kazanas is conformally equivalent to the Schwarzschild--De Sitter spacetime~\cite{footnote,Mokdad:2017udm},
and the only true physical singularity is at $r=0$. 
The existence of additional horizons modifies the causal structure of the spacetime, but the spatial topology outside of any event horizon remains that of $\mathbb{R}^3\smallsetminus \{ 0 \}$ (i.e., a point-like singularity removed
from a space-like section), as in the Schwarzschild case. The Mannheim--Kazanas spacetime
is thus homotopologically equivalent to $\mathbb{R} \times (\mathbb{R}^3\smallsetminus \{0\})$. By employing that $\mathbb{R}^3\smallsetminus \{ 0 \} \cong S^2 \times \mathbb{R}^+$, we get that the Mannheim--Kazanas spacetime is homotopic to $\mathbb{R} \times S^2 \times \mathbb{R}^+$. The cross product theorem from homotopy theory~\cite{Nakahara} ensures that the corresponding fundamental group (or the first homotopy group) for this spacetime is $\pi_1(\mathbb{R} \times  S^2 \times \mathbb{R}^+) \cong \pi_1(\mathbb{R}) \otimes  \pi_1(S^2) \otimes \pi_1(\mathbb{R}^+) \cong 1$ ($\otimes$ denotes a group product). The latter means that all closed loops on the manifold can be contracted smoothly to a point, so $\pi_1$ has only one element (only the basic class of curves) --- the identity element, so that
the manifold is homotopically trivial.
First non-trivial homotopy group is the second one which is   $\pi_2(\mathbb{R} \times  S^2 \times \mathbb{R}^+) \cong \pi_2(S^2) \cong \mathbb{Z}$.  All higher homotopy groups are non-trivial since $\pi_{k>2}(S^2)$ are non-trivial (though not well known explicitly for too large $k$).}







{\cdblue Let us now consider the wormholes~(\ref{66.dd}) that connect {two asymptotically identical universes}. Any such asymptotic spacetime is topologically 
$\mathbb{R} \times S^2 \times \mathbb{R}^+$ with the boundary $\mathbb{R} \times S^2$.  The wormholes thus correspond to two  $\mathbb{R} \times S^2 \times \mathbb{R}^+$ manifolds glued together at $\mathbb{R} \times S^2$ --- so the wormholes are topologically $S^2\times \mathbb{R}^2$, which means that the resulting fundamental group $\pi_1(S^2\times \mathbb{R}^2)\cong \pi_1(S^2)\otimes \pi_1(\mathbb{R}^2)\cong 1$ is trivial.
Similar to the Mannheim--Kazanas case, the second homotopy group $\pi_2(S^2\times \mathbb{R}^2)\cong \pi_2(S^2)\otimes \pi_2(\mathbb{R}^2) \cong \mathbb{Z}$, is non-trivial along with other higher homotopy groups. So in this case the spherically symmetric WCG wormholes~(\ref{66.dd}) are homotopically equivalent to the Mannheim--Kazanas black holes. }


{\cdblue The situation is, however, different when the wormholes~(\ref{66.dd}) connect {the same universe with itself}. These single-universe wormholes can still be described by the wormhole metric~(\ref{V.68.kl}), since the metric does not determine the topology of spacetime, and locally the manifold of a wormhole connecting two spacetimes or one spacetime can clearly have the same metric form. To obtain the single-universe wormholes
we can, for instance, glue together boundaries of the two separate universes from the previous paragraph at some large $|r|$. Since both universes are, according to~(\ref{66.dd}), parity images of each other, such a gluing could be made smooth while still locally being a solution of the Bach vacuum equation.  Note that locally, each spatial slice looks like $S^2 \times \mathbb{R}^+$, but globally, the presence of the wormhole introduces
non-contractible loops in spacetime, leading to a non-trivial fundamental group $\pi_1$. 
For example, if a test particle (e.g. photon) travels from one region of space, passes through the wormhole, and returns to the original region via another path, the corresponding trajectory is non-contractible within that universe. Since $\pi_1$ is non-trivial, the single-universe WCG wormholes are homotopically distinct from the Mannheim--Kazanas black holes.

It is also worth noting that while a Weyl transformation affects the geometric quantities of a manifold, it does not affect its topological properties. Thus, a homotopy classification valid in one spacetime is valid in all Weyl-equivalent spacetimes within a given WCG. Consequently, the above homotopy arguments imply that the single-universe WCG wormholes and the Mannheim-Kazanas black holes represent inequivalent spacetime configurations in WCG.

While the above homotopy argument could not distinguish topologically between the Mannheim--Kazanas solution and the two-universe WCG wormholes, other topological invariants like Betti numbers can show such a distinction.   Betti numbers describe the structure of a topological space by counting the number of independent ``holes'' of each dimension, from a strict mathematical point of view the $n$-th Betti number represents the rank of the $n$-th homology group~\cite{Nakahara}.
In this setup, we  examine Betti numbers 
of the two respective spacetimes by focusing on the topology of the spatial slices.
The zeroth Betti number $B_0$ counts the number of connected components, which is $1$ for both the Mannheim--Kazanas and the WCG wormhole cases. 
The first Betti number $B_1$ counts the number of $1$-dimensional loops, such as closed curves. In both the Mannheim--Kazanas case and the WCG wormhole case there are no non-trivial loops, so for both spacetimes $B_1 =0$. The second Betti number $B_2$ counts the number of independent $2$-dimensional surfaces. While in the Mannheim--Kazanas spacetimes there are no non-trivial $2$-dimensional ``holes'' (hence $B_2 = 0$), in the WCG wormhole spatial slice there is one $2$-dimensional ``hole'' corresponding to the $S^2$ cross-section of the throat (hence $B_2 = 1$). Higher Betti numbers $B_{k>2}$ are in both cases zero as there are no higher-dimensional voids in spatial slices.
So, finally for the Mannheim--Kazanas spacetimes we have 
\begin{eqnarray}
(B_0,B_1,B_2,B_{k>2}) \ = \ (1,0,0,0)\, ,
\end{eqnarray}
while for the two-universe WCG wormholes that connect two asymptotically identical universes 
\begin{eqnarray}
(B_0,B_1,B_2,B_{k>2}) \ = \ (1,0,1,0)\, .
\end{eqnarray}
In a manner analogous to homotopy classes, the homology groups and ensuing Betti numbers are topological invariants that remain constant under a Weyl transformation, implying that both aforementioned spacetimes are Weyl inequivalent.}



%

\section{Discussion and Conclusions}
\label{Concl.}

In this paper, we have used the Newman--Penrose tetrad formalism, known from Einstein's general relativity, to analyze the external solution for a static, spherically symmetric source (with and without electromagnetic field) in Weyl conformal gravity.  First, we have demonstrated the efficiency of the NP formalism by deriving some of the well-known results in WCG in a new and shorter way. In Einstein's general relativity, the vacuum field equations imply that all NP-Ricci scalars are zero. Substituting these back into the NP field equations yields a coupled system of first-order differential equations.  We have shown that a similar situation holds in WCG, where the NP formalism allows us to trade the fourth-order Bach equations for a system of at most second-order equations. This, in turn, may allow some of the field equations to be solved more easily.  To illustrate this, we have focused primarily on external solutions for static, spherically symmetric sources.  In a second step, we have formulated a unifying framework  for solving the electrovacuum WCG equations in the NP formalism and in arbitrary gauge. Apart from already known solutions in WCG, such as the Mannheim--Kazanas black hole solution and the charged Reissner--Nordstr\"{o}m-like solution (by including the electromagnetic field in the Bach equation), we also derived a two-parameter class of traversable WCG wormhole solutions, which do not exist in Einstein gravity, although for a certain choice of parameters the vicinity of the wormhole throat can be identified with the throat region of the Ellis--Bronnikov wormhole.

All considered solutions corresponded to spacetimes of Petrov ``type-D'' character. Since Petrov's algebraic classification of space-times is based on properties of the Weyl tensor, one might suspect that the reason why the NP formalism is so peculiarly well suited to the study of black hole solutions in WCG is that some analogue of the Goldberg--Sachs theorem must hold in WCG.


In order to present the essential features as simply as possible, we did not go beyond static, spherically symmetric sources in our exposition. For axially symmetric solutions, such as the Kerr and Kerr--Newman types of WCG solutions, the NP formalism is also instrumental in simplifying the solutions. This can be demonstrated by coupling the NP formalism with the Newman--Janis algorithm, and will be discussed in our follow-up paper. {\cdblue Similarly, in our future work we plan to discuss Petrov ``type N'' spacetimes (such as those associated with longitudinal gravitational radiation and the peeling theorem), since their Bach equations have a particularly simple structure in the NP formalism.  }

\medskip

\begin{acknowledgments}
We thank Teresa Lehe\v{c}kov\'{a} for useful discussions and comments on the manuscript. 
P.J. was in part supported by the FNSPE CTU grant RVO14000. 
\end{acknowledgments}

\vspace{1cm}
\appendix

\section{NP spin-coefficient equations~\label{Appendix A0}}

The Weyl and Ricci scalars can be calculated from the spin coefficients by the following set of equations. These are derived from known identities in differential geometry and are therefore valid in all theories.
Here we list some of the requisite formulas needed in the text. Explicit derivations (modulo sign convention) can be found, for example, in Ref.~\cite{Chandra}.
\begin{widetext}
\begin{eqnarray}
    &&\mathrm{D}\sigma \ - \ \delta\kappa  \  = \ \sigma\left(3\epsilon \ - \ \bar{\epsilon} \ + \ \rho \ + \ \bar{\rho}\right) \ + \ \kappa\left(\bar{\pi} \ - \ \tau \ - \ 3\beta \ - \ \alpha\right) \ + \ \Psi_0\, ,  \label{eqNPForWeyl1} \nonumber \\[2mm]
    &&\mathrm{D} \rho \ - \ \bar{\delta} \kappa \ = \ \left(\rho^2 \ + \ \sigma \bar{\sigma}\right) \ + \ \rho(\epsilon \ + \ \bar{\epsilon}) \ - \ \bar{\kappa} \tau \ + \ \kappa(\pi \ - \ 3 \alpha \ - \ \bar{\beta}) \ + \ \Phi_{00}\, , \nonumber \\[2mm]
    &&\mathrm{D} \tau \ - \ \Delta \kappa \ = \ \rho(\tau\ + \ \bar{\pi}) \ + \ \sigma(\bar{\tau} \ + \ \pi) \ + \ \tau(\epsilon \ - \ \bar{\epsilon}) \ - \ \kappa(3 \gamma \ + \ \bar{\gamma}) \ + \ \Psi_1+\Phi_{01}\, , \nonumber \\[2mm]
    &&
    \mathrm{D} \alpha \ - \ \bar{\delta} \epsilon \ = \ \alpha(\rho \ + \ \bar{\epsilon} \ - \ 2 \epsilon) \ + \ \beta \bar{\sigma} \ - \ \bar{\beta} \epsilon \ - \ \kappa \lambda \ - \ \bar{\kappa} \gamma \ + \ \pi(\epsilon \ + \ \rho) \ + \ \Phi_{10}\, , \nonumber \\[2mm]
    && \mathrm{D} \beta \ - \ \delta \epsilon  \ = \ \sigma(\alpha \ + \ \pi) \ + \ \beta(\bar{\rho} \ - \ \bar{\epsilon}) \ - \ \kappa(\mu \ + \ \gamma) \ + \ \epsilon(\bar{\pi} \ - \ \bar{\alpha}) \ + \ \Psi_1\, , \label{eqNPForWeyl2} \nonumber \\[2mm]
    && \mathrm{D} \lambda \ - \ \bar{\delta} \pi  \ = \ (\rho \lambda \ + \ \bar{\sigma} \mu) \ + \ \pi(\pi \ + \ \alpha \ - \ \beta) \ - \ \nu \bar{\kappa} \ + \ \lambda(\bar{\epsilon}\ - \ 3 \epsilon) \ + \ \Phi_{20}\, , \nonumber \\[2mm]
    && \mathrm{D} \nu \ - \ \Delta \pi \  = \ \mu(\pi \ + \ \bar{\tau}) \ + \ \lambda(\bar{\pi} \ + \ \tau) \ + \ \pi(\gamma \ - \ \bar{\gamma})\ - \ \nu(3 \epsilon \ + \ \bar{\epsilon})\ + \ \Psi_3 \ + \ \Phi_{21}\, , \label{eqNPForWeyl3}\nonumber \\[2mm]
    && \Delta \alpha \ - \ \bar{\delta} \gamma \ = \ \nu(\rho \ + \ \epsilon) \ - \ \lambda(\tau \ + \ \beta) \ + \ \alpha(\bar{\gamma} \ - \ \bar{\mu}) \ + \ \gamma(\bar{\beta} \ - \ \bar{\tau}) \ - \ \Psi_3\, , \nonumber \\[2mm]
    &&\Delta \lambda \ - \ \bar{\delta} \nu \  = \ \lambda(\bar{\gamma} \  - \ 3 \gamma \ - \ \mu \ - \ \bar{\mu}) \ + \ \nu(3 \alpha \ + \ \bar{\beta} \ + \ \pi \ - \ \bar{\tau}) \ - \ \Psi_4,  \label{eqNPForWeyl4}  \nonumber \\[2mm] 
    &&\delta \rho \ - \ \bar{\delta} \sigma \ = \ \rho(\bar{\alpha} \ + \ \beta) \ + \ \sigma(\bar{\beta} \ - \ 3 \alpha) \ + \ \tau(\rho \ - \ \bar{\rho}) \ + \ \kappa(\mu \ - \ \bar{\mu}) \ - \ \Psi_1 \ + \ \Phi_{01}\, , \nonumber \\[2mm] 
    &&\delta \lambda \ - \ \bar{\delta} \mu \ = \ \nu(\rho \ - \ \bar{\rho}) \ + \ \pi(\mu \ - \ \bar{\mu}) \ + \ \mu(\alpha \ + \ \bar{\beta}) \ + \ \lambda(\bar{\alpha}\ - \ 3 \beta) \ - \ \Psi_3 \ + \ \Phi_{21}\, , \nonumber \\[2mm] &&\delta \nu \ - \ \Delta \mu \ = \ \left(\mu^2 \ + \ \lambda \bar{\lambda}\right) \ + \ \mu(\gamma \ + \ \bar{\gamma}) \ - \ \bar{\nu} \pi \ + \ \nu(\tau \ - \ 3 \beta \ - \ \bar{\alpha}) \ + \ \Phi_{22}\, , \nonumber \\[2mm] &&\delta \gamma \ - \ \Delta \beta \ = \ \gamma(\tau \ - \ \bar{\alpha} \ - \ \beta)\ + \ \mu \tau \ - \ \sigma \nu \ - \ \epsilon \bar{\nu} \ + \ \beta(\mu \ - \ \gamma \ + \ \bar{\gamma}) \ + \ \alpha \bar{\lambda} \ + \ \Phi_{12}\, , \nonumber \\[2mm] 
    &&\delta \tau \ - \ \Delta \sigma \ = \ ( \mu \sigma \ + \ \bar{\lambda} \rho) \ + \ \tau(\tau \ + \ \beta \ - \ \bar{\alpha}) \ + \ \sigma(\bar{\gamma} \ - \ 3 \gamma) \ - \ \kappa \bar{\nu} \ + \ \Phi_{02}\, , \nonumber \\[2mm] 
    && \mathrm{D} \mu \ - \ \delta \pi \ = \ (\bar{\rho} \mu \ + \ \sigma \lambda) \ + \ \pi(\bar{\pi} \ - \ \bar{\alpha} \ + \ \beta) \ - \ \mu(\epsilon \ + \ \bar{\epsilon}) \ - \ \nu \kappa \ + \ \Psi_2 \ + \ 2\Lambda\, , \nonumber \\[2mm] 
    &&\Delta \rho \ - \ \bar{\delta} \tau \ = \ -(\rho \bar{\mu} \ + \ \sigma \lambda) \ + \ \tau(\bar{\beta} \ - \ \alpha \ - \ \bar{\tau}) \ + \ \rho(\gamma \ + \ \bar{\gamma}) \ + \ \nu \kappa \ - \ \Psi_2 \ - \ 2\Lambda\, , \label{eqNPForWeyl12}\nonumber \\[2mm] 
    &&\mathrm{D} \gamma\ - \ \Delta \epsilon \  = \ \alpha(\tau \ + \ \bar{\pi}) \ + \ \beta(\bar{\tau} \ + \ \pi) \ - \ \gamma(\epsilon \ + \ \bar{\epsilon}) \ - \ \epsilon(\gamma \ + \ \bar{\gamma}) \ + \ \tau \pi \ - \ \nu \kappa \ + \ \Psi_2 \ + \ \Phi_{11} \ - \ \Lambda\, , \label{eqNPForWeyl10}\nonumber \\[2mm] 
    &&\delta \alpha \ - \ \bar{\delta} \beta \  = \
     (\mu \rho-\lambda \sigma) \ + \ \alpha \bar{\alpha} \ + \ \beta \bar{\beta} \ - \ 2 \alpha \beta \ + \ \gamma(\rho \ - \ \bar{\rho}) \ + \ \epsilon(\mu \ - \ \bar{\mu}) \ - \ \Psi_2 \ + \ \Phi_{11} \ + \ \Lambda\, .
    \label{eqNPForWeyl11}
\end{eqnarray}
\end{widetext}

\section{The Bach
equations in the NP formalism \label{Appendix A}}


In this Appendix, we list the requisite equations that are 
relevant to Secs.~\ref{MK}, \ref{Sec.III.C.cv}, and \ref{Sec.V.cd} and to the subsequent work.
%
First, we list the components of the Bach tensor in NP formalism. In the following formulas ``c.c.'' stands for the complex conjugate part. For more technical details the reader can consult e.g. Ref.~\cite{Chandra,Sv}. 

From~(\ref{Z42A}), we know that the Bach tensor can be written as a difference of two terms: the fourth derivative part and the Ricci tensor part. We start with the fourth derivative part $B^{(1)}_{\mu\nu}=\nabla^\alpha \nabla^\beta \tensor{C}{_\mu_\alpha_\nu_\beta}$. In the NP formalism it is given by~\cite{Sv}
\begin{widetext}
\label{eq:NPBach}
\begin{eqnarray}
\label{eq:BachNP00}
B_{ll}^{(1)} &=& \bar{\delta} \Psi_0-\mathrm{D} \bar{\delta} \Psi_1 \ - \ \bar{\delta} \mathrm{D} \Psi_1 \ + \ \mathrm{DD} \Psi_2 \ + \ \lambda \mathrm{D} \Psi_0 \ + \ \bar{\sigma} \Delta \Psi_0 \ + \ (2 \pi \ - \ 7 \alpha \ - \ \bar{\beta}) \bar{\delta} \Psi_0\nonumber  \\[2mm]
&& + \ (5 \alpha \ + \ \bar{\beta} \ - \ 3 \pi) \mathrm{D} \Psi_1 \ - \ \bar{\kappa} \Delta \Psi_1 \ - \ \bar{\sigma} \delta \Psi_1 \ + \ (3 \epsilon \ + \ \bar{\epsilon} \ + \ 7 \rho) \bar{\delta} \Psi_1 \nonumber \\[2mm]
&& - \ (\epsilon\ + \ \bar{\epsilon} \ + \ 6 \rho) \mathrm{D} \Psi_2 \ + \ \bar{\kappa} \delta \Psi_2 \ - \ 5 \kappa \bar{\delta} \Psi_2 \ + \ 4 \kappa \mathrm{D} \Psi_3 \nonumber \\[2mm]
&& + \ \Psi_0[\bar{\kappa} \nu \ + \ 4 \alpha(3 \alpha \ + \ \bar{\beta}) \ - \ (\epsilon \ + \ \bar{\epsilon} \ + \ 3 \rho) \lambda \ + \ \pi(\pi \ - \ 7 \alpha \ - \ \bar{\beta})\ + \ \bar{\sigma}(\mu \ - \ 4 \gamma) \nonumber \\[2mm]
&&+ \ \mathrm{D} \lambda \ - \ 4 \bar{\delta} \alpha \ + \ \bar{\delta} \pi] \nonumber \\[2mm]
&& + \ 2 \Psi_1[2 \kappa \lambda \ + \ \bar{\kappa}(\gamma-\mu) \ + \ \rho(5 \pi \ - \ 9 \alpha \ - \ 2 \bar{\beta}) \ + \ \bar{\sigma}(\beta \ + \ 2 \tau) \ + \ \epsilon(2 \pi \ - \ 4 \alpha \ - \ \bar{\beta}) \nonumber \\[2mm]
&& + \ \bar{\epsilon}(\pi \ - \ \alpha) \ + \ \mathrm{D} \alpha \ - \ \mathrm{D} \pi \ + \ \bar{\delta} \epsilon \ + \ 2 \bar{\delta} \rho] \nonumber \\[2mm]
&& + \ 3 \Psi_2[\kappa(3 \alpha \ + \ \bar{\beta} \ - \ 3 \pi) \ - \ \bar{\kappa} \tau \ + \ \rho(\epsilon \ + \  \bar{\epsilon}+3 \rho)\ - \ \sigma \bar{\sigma} \ - \ \mathrm{D} \rho \ - \ \bar{\delta} \kappa] \nonumber \\[2mm]
&&  + \ 2 \Psi_3[\kappa(\epsilon \ - \ \bar{\epsilon} \ - \ 5 \rho) \ + \  \bar{\kappa} \sigma \ + \ \mathrm{D} \kappa] \ + \ 2 \Psi_4 \kappa^2 \ + \ \mbox{c.c.}\, .
\end{eqnarray}

\begin{eqnarray}
B_{lm}^{(1)}&=&\bar{\delta} \Delta \Psi_0 \ - \ \mathrm{D} \Delta \Psi_1 \ - \  \bar{\delta} \delta \Psi_1 \ + \ \mathrm{D} \delta \Psi_2 \nonumber \\[2mm]
&&  + \ \nu \mathrm{D} \Psi_0 \ + \ (\pi \ - \ 3 \alpha \ + \ \bar{\beta}) \Delta \Psi_0 \ + \ (\mu \ - \ \bar{\mu} \ - \ 4 \gamma) \bar{\delta} \Psi_0 \nonumber \\[2mm]
&& + \ (2 \gamma \ - \ 2 \mu \ + \ \bar{\mu}) \mathrm{D} \Psi_1 \ + \ (\epsilon \ - \ \bar{\epsilon} \ + \ 3 \rho) \Delta \Psi_1 \ + \ (3 \alpha \ - \ \bar{\beta} \ - \ \pi) \delta \Psi_1\nonumber \\[2mm]
&& + \ (2 \beta \ + \ \bar{\pi} \ + \ 4 \tau) \bar{\delta} \Psi_1 \nonumber \\[2mm]
&&  - \ (\bar{\pi} \ + \ 3 \tau) \mathrm{D} \Psi_2 \ - \ 2 \kappa \Delta \Psi_2 \ - \ (\epsilon \ - \ \bar{\epsilon} \ + \ 3 \rho) \delta \Psi_2 \ - \ 3 \sigma \bar{\delta} \Psi_2 \ + \ 2 \sigma \mathrm{D} \Psi_3 \ + \ 2 \kappa \delta \Psi_3 \nonumber \\[2mm]
&&  + \ \Psi_0[(4 \gamma \ - \ \mu)(3 \alpha \ - \ \bar{\beta} \  - \ \pi) \ + \ \bar{\mu}(4 \alpha \ - \ \pi) \ + \ \nu(\bar{\epsilon}\ - \ \epsilon \ - \ 3 \rho) \ - \ \lambda \bar{\pi}\nonumber \\[2mm]
&&  + \ \mathrm{D} \nu \ - \ 4 \bar{\delta} \gamma \ + \ \bar{\delta} \mu]\nonumber \\[2mm]
&&  + \ 2 \Psi_1[2 \kappa \nu \ + \ (\mu \ - \ \gamma)(\epsilon \ - \ \bar{\epsilon} \ + \ 3 \rho) \ - \ \bar{\mu}(2 \rho \ + \ \epsilon) \ + \ (\beta \ + \ 2 \tau)(\pi \ - \ 3 \alpha \ + \ \bar{\beta}) \nonumber \\[2mm]
&&  + \ \bar{\pi}(\pi \ - \ \alpha) \ + \ \mathrm{D} \gamma \ - \ \mathrm{D} \mu \ + \ \bar{\delta} \beta \ + \ 2 \bar{\delta} \tau] \nonumber \\[2mm]
&&  + \ 3 \Psi_2[\kappa(\bar{\mu} \ - \ 2 \mu) \ + \ \bar{\pi} \rho \ + \ \sigma(3 \alpha \ - \ \bar{\beta} \ - \ \pi) \ + \ \tau(\epsilon \ - \ \bar{\epsilon} \ + \ 3 \rho) \ - \ \mathrm{D} \tau \ - \ \bar{\delta} \sigma] \nonumber \\[2mm]
&& + \ 2 \Psi_3[\kappa(2 \beta \ - \ \bar{\pi} \ - \ 2 \tau) \ + \ \sigma(\bar{\epsilon} \ - \ \epsilon \ - \ 3 \rho) \ + \ \mathrm{D} \sigma] \ + \ 2 \Psi_4 \kappa \sigma \nonumber \\[2mm]
&& + \ \delta \delta \bar{\Psi}_1 \ - \ \delta \mathrm{D} \bar{\Psi}_2 \ - \ \mathrm{D} \delta \bar{\Psi}_2 \ + \ \mathrm{DD} \bar{\Psi}_3 \nonumber \\[2mm]
&& - \ 2 \bar{\lambda} \delta \bar{\Psi}_0 \ + \ 3 \bar{\lambda} \mathrm{D} \bar{\Psi}_1 \ + \ \sigma \Delta \bar{\Psi}_1 \ + \ (4 \bar{\pi} \ - \ 3 \bar{\alpha} \ - \ \beta) \delta \bar{\Psi}_1 \nonumber \\[2mm]
&&  + \ (\bar{\alpha} \ + \ \beta \ - \ 5 \bar{\pi}) \mathrm{D} \bar{\Psi}_2 \ - \ \kappa \Delta \bar{\Psi}_2 \ + \ (\epsilon \ - \ \bar{\epsilon}\ + \ 5 \bar{\rho}) \delta \bar{\Psi}_2 \ - \ \sigma \bar{\delta} \bar{\Psi}_2 \nonumber \\[2mm]
&& + \ (3 \bar{\epsilon} \ - \ \epsilon \ - \ 4 \bar{\rho}) \mathrm{D} \bar{\Psi}_3 \ - \ 3 \bar{\kappa} \delta \bar{\Psi}_3 \ + \ \kappa \bar{\delta} \bar{\Psi}_3 \ + \ 2 \bar{\kappa} \mathrm{D} \bar{\Psi}_4 \nonumber \\[2mm]
&&  + \ \bar{\Psi}_0[\bar{\lambda}(5 \bar{\alpha} \ + \ \beta \ - \ 3 \bar{\pi}) \ - \ \bar{\nu} \sigma \ - \ \delta \bar{\lambda}] \nonumber \\[2mm]
&& + \ 2 \bar{\Psi}_1[\kappa \bar{\nu} \ + \ \bar{\alpha}(\bar{\alpha} \ + \ \beta) \ + \ \bar{\pi}(2 \bar{\pi} \ - \ 3 \bar{\alpha} \ - \ \beta) \ - \ \bar{\lambda}(4 \bar{\rho} \ + \ \epsilon) \ + \ \sigma(\bar{\mu} \ - \ \bar{\gamma}) \nonumber \\[2mm]
&&  + \ \mathrm{D} \bar{\lambda} \ - \ \delta \bar{\alpha} \ + \ \delta \bar{\pi}] \nonumber \\[2mm]
&&  + \ 3 \bar{\Psi}_2[2 \bar{\kappa} \bar{\lambda} \ - \ \kappa \bar{\mu} \ + \ \bar{\pi}(\epsilon \ - \ \bar{\epsilon}) \ + \ \bar{\rho}(4 \bar{\pi} \ - \  \bar{\alpha} \ - \ \beta) \ + \ \sigma \bar{\tau} \ - \ \mathrm{D} \bar{\pi} \ + \ \delta \bar{\rho}] \nonumber \\[2mm]
&&  + \ 2 \bar{\Psi}_3(\kappa(\bar{\beta} \ - \ \bar{\tau}) \ + \ \bar{\kappa}(\beta \ - \ 4 \bar{\pi}) \ - \ \sigma \bar{\sigma} \ + \ (\bar{\rho} \ - \ \bar{\epsilon})(\epsilon \ - \ \bar{\epsilon} \ + \ 2 \bar{\rho}) \ + \ \mathrm{D} \bar{\epsilon} \ -\ \mathrm{D} \bar{\rho} \ - \ \delta \bar{\kappa}) \nonumber \\[2mm]
&& + \ \bar{\Psi}_4[\bar{\kappa}(5 \bar{\epsilon} \ - \ \epsilon \ - \  3 \bar{\rho}) \ + \ \kappa \bar{\sigma} \ + \ \mathrm{D} \bar{\kappa}]\, , 
\end{eqnarray}

\begin{eqnarray}
B_{ln}^{(1)} &=& \bar{\delta} \Delta \Psi_1 \ - \ \mathrm{D} \Delta \Psi_2 \ - \ \bar{\delta} \delta \Psi_2 \ + \ \mathrm{D} \delta \Psi_3 \ - \ \lambda \Delta \Psi_0-\nu \bar{\delta} \Psi_0 \nonumber \\[2mm]
&& + \  2 \nu \mathrm{D} \Psi_1 \ + \ (2 \pi \ - \ \alpha \ + \  \bar{\beta}) \Delta \Psi_1 \ + \  \lambda \delta \Psi_1 \ + \ (2 \mu \ - \ \bar{\mu} \ - \ 2 \gamma) \bar{\delta} \Psi_1 \nonumber \\[2mm]
&& + \ (\bar{\mu} \ - \ 3 \mu) \mathrm{D} \Psi_2 \ +  \ (2 \rho \ - \ \epsilon \ - \ \bar{\epsilon}) \Delta \Psi_2 \ + \ (\alpha \ - \ \bar{\beta} \ - \ 2 \pi) \delta \Psi_2 \ + \ (\bar{\pi} \ + \ 3 \tau) \bar{\delta} \Psi_2 \nonumber \\[2mm]
&& + \ (2 \beta \ - \ \bar{\pi} \ - \ 2 \tau) \mathrm{D} \Psi_3 \ - \ \kappa \Delta \Psi_3 \ + \ (\epsilon \ + \ \bar{\epsilon} \ - \ 2 \rho) \delta \Psi_3 \ -  \ 2 \sigma \bar{\delta} \Psi_3 \ + \  \sigma \mathrm{D} \Psi_4 \ +  \ \kappa \delta \Psi_4 \nonumber \\[2mm]
&& + \ \Psi_0[\lambda(4 \gamma \ - \ \mu \ + \ \bar{\mu}) \ + \ \nu(\alpha \ - \ \bar{\beta} \ - \ 2 \pi)\ - \ \bar{\delta} \nu] \nonumber \\[2mm]
&& + \ 2 \Psi_1[\gamma(\alpha \ - \ \bar{\beta} \ - \ 2 \pi) \ - \ \lambda(\beta \ + \ \bar{\pi} \ + \  2 \tau)\ + \ \mu(\bar{\beta}\ - \ \alpha \ + \ 2 \pi) \ + \ \bar{\mu}(\alpha\ - \ \pi) \nonumber \\[2mm]
&& + \ \nu(\epsilon \ + \ \bar{\epsilon} \ - \ 2 \rho) \ + \ \mathrm{D} \nu \ - \ \bar{\delta} \gamma \ + \ \bar{\delta} \mu] \nonumber \\[2mm]
&& + \ 3 \Psi_2[\kappa \nu \ + \ \mu(2 \rho \ - \ \epsilon \ - \ \bar{\epsilon}) \ - \ \bar{\mu} \rho \ + \ \pi \bar{\pi} \ + \ \lambda \sigma \ + \ \tau(2 \pi \ - \ \alpha  \ + \ \bar{\beta}) \ - \ \mathrm{D} \mu \ + \ \bar{\delta} \tau] \nonumber \\[2mm]
&& + \ 2 \Psi_3[\kappa(\bar{\mu} \ - \ 2 \mu \ - \ \gamma) \ + \ \epsilon(\beta \ - \ \tau \ - \ \bar{\pi}) \ + \ \bar{\epsilon}(\beta \ - \ \tau) \ + \ \rho(\bar{\pi} \ - \ 2 \beta \ + \ 2 \tau) \nonumber \\[2mm]
&& + \ \sigma(\alpha \ - \ \bar{\beta} \ - \ 2 \pi) \ + \ \mathrm{D} \beta \ - \ \mathrm{D} \tau \ - \ \bar{\delta} \sigma] \nonumber \\[2mm]
&& + \ \Psi_4[\kappa(4 \beta \ - \ \bar{\pi} \ - \ \tau) \ + \  \sigma(\epsilon \ + \ \bar{\epsilon} \ - \ 2 \rho)\ + \ \mathrm{D} \sigma] \ + \ \mbox{c.c.}\, .
\end{eqnarray}

\begin{eqnarray}
B_{nm}^{(1)}&=& \Delta \Delta \Psi_1 \ - \ \Delta \delta \Psi_2 \ - \ \delta \Delta \Psi_2 \ + \ \delta \delta \Psi_3 \nonumber \\[2mm]
&&  - \ 2 \nu \Delta \Psi_0 \ + \ (4 \mu \ - \ 3 \gamma \ + \ \bar{\gamma}) \Delta \Psi_1 \  + \ 3 \nu \delta \Psi_1 \ - \ \bar{\nu} \bar{\delta} \Psi_1 \nonumber \\[2mm]
&&  + \ \bar{\nu} \mathrm{D} \Psi_2 \ + \ (5 \tau \ - \ \bar{\alpha} \ - \ \beta) \Delta \Psi_2 \ + \ (\gamma \ - \ \bar{\gamma} \ - \ 5 \mu) \delta \Psi_2 \ + \ \bar{\lambda} \bar{\delta} \Psi_2 \nonumber \\[2mm]
&&  - \ \bar{\lambda} \mathrm{D} \Psi_3 \ - \ 3 \sigma \Delta \Psi_3 \ + \ (\bar{\alpha} \ + \ 3 \beta \ - \ 4 \tau) \delta \Psi_3 \ + \ 2 \sigma \delta \Psi_4 \nonumber \\[2mm]
&&  + \ \Psi_0[\nu(5 \gamma \ - \ \bar{\gamma} \ - \ 3 \mu) \ + \ \lambda \bar{\nu} \ - \ \Delta \nu] \nonumber \\[2mm]
&&  + \ 2 \Psi_1[\nu(\bar{\alpha} \ - \ 4 \tau) \ + \ \bar{\nu}(\alpha \ - \ \pi) \ - \ \lambda \bar{\lambda} \ + \ (\gamma \ - \ \mu)(\gamma \ - \ \bar{\gamma} \ - \ 2 \mu) \nonumber \\[2mm]
&&  - \ \Delta \gamma \ + \ \Delta \mu \ + \ \delta \nu] \nonumber \\[2mm]
&&  + \ 3 \Psi_2[\mu(4 \tau \ - \ \bar{\alpha} \ - \ \beta) \ + \ \bar{\lambda} \pi \ - \ \bar{\nu} \rho \ + \ 2 \nu \sigma \ + \ \tau(\bar{\gamma} \ - \ \gamma) \ + \ \Delta \tau \ - \ \delta \mu] \nonumber \\[2mm]
&&  + \ 2 \Psi_3[\kappa \bar{\nu} \ - \ \sigma(\bar{\gamma} \ + \ 4 \mu) \ + \ \tau(2 \tau \ - \ \bar{\alpha} \ - \ 3 \beta) \ + \ \beta(\bar{\alpha} \ + \ \beta)\nonumber \\[2mm]
&& + \ \bar{\lambda}(\rho \ - \ \epsilon) \ - \ \Delta \sigma \ + \ \delta \beta \ - \ \delta \tau] \nonumber \\[2mm]
&&  + \ \Psi_4[ \ - \ \kappa \bar{\lambda} \ + \ \sigma(\bar{\alpha} \ + \ 5 \beta \ - \ 3 \tau) \ + \ \delta \sigma] \nonumber \\[2mm]
&&  - \ \Delta \mathrm{D} \bar{\Psi}_3 \ + \ \Delta \delta \bar{\Psi}_2 \ + \ \bar{\delta} \mathrm{D} \bar{\Psi}_4 \ - \ \bar{\delta} \delta \bar{\Psi}_3 \nonumber \\[2mm]
&&  - \ 2 \bar{\lambda} \Delta \bar{\Psi}_1 \ - \ 2 \bar{\nu} \delta \bar{\Psi}_1 \ + \ 2 \bar{\nu} \mathrm{D} \bar{\Psi}_2 \ + \ (3 \bar{\pi} \ + \ \tau) \Delta \bar{\Psi}_2 \ + \ (\bar{\gamma} \ - \ \gamma \ + \ 3 \bar{\mu}) \delta \bar{\Psi}_2 \ + \ 3 \bar{\lambda} \bar{\delta} \bar{\Psi}_2 \nonumber \\[2mm]
&&   + \ (\gamma \ - \ \bar{\gamma} \ - \ 3 \bar{\mu}) \mathrm{D} \bar{\Psi}_3 \ + \ (2 \bar{\rho} \ - \ \rho \ - \ 2 \bar{\epsilon}) \Delta \bar{\Psi}_3 \ + \ (\alpha \ - \ 3 \bar{\beta} \ + \ \bar{\tau}) \delta \bar{\Psi}_3 \nonumber \\[2mm]
&& - \ (2 \bar{\alpha} \ + \ 4 \bar{\pi} \ + \ \tau) \bar{\delta} \bar{\Psi}_3 \ + \ (3 \bar{\beta} \ - \ \alpha \ - \ \bar{\tau}) \mathrm{D} \bar{\Psi}_4 \ - \ \bar{\kappa} \Delta \bar{\Psi}_4 \ + \ (4 \bar{\epsilon} \ +  \ \rho \ - \ \bar{\rho}) \bar{\delta} \bar{\Psi}_4 \nonumber \\[2mm]
&&  + \ 2 \bar{\Psi}_0 \bar{\lambda} \bar{\nu} \ + \ 2 \bar{\Psi}_1[\bar{\lambda}(\gamma \ - \ \bar{\gamma} \ - \ 3 \bar{\mu}) \ + \ \bar{\nu}(2 \bar{\alpha} \ - \ 2 \bar{\pi} \ - \ \tau) \ - \ \Delta \bar{\lambda}] \nonumber \\[2mm]
&&  + \ 3 \bar{\Psi}_2[\bar{\lambda}(3 \bar{\beta} \ - \ \bar{\tau} \ - \ \alpha) \ + \ \bar{\pi}(3 \bar{\mu} \ - \ \gamma \ + \ \bar{\gamma}) \ + \ \bar{\nu}(\rho \ - \ 2 \bar{\rho}) \ + \ \bar{\mu} \tau \ + \ \Delta \bar{\pi} \ + \ \bar{\delta} \bar{\lambda}] \nonumber \\[2mm]
&&  + \ 2 \bar{\Psi}_3[2 \bar{\kappa} \bar{\nu} \ + \ (\bar{\epsilon} \ -  \ \bar{\rho})(\gamma \ - \ \bar{\gamma} \ - \ 3 \bar{\mu}) \ - \ \rho(\bar{\gamma} \ + \ 2 \bar{\mu}) \ + \ \tau(\bar{\tau} \ - \ \bar{\beta})   \nonumber \\[2mm]
&&  + \ (\bar{\alpha} \ + \ 2 \bar{\pi})(\alpha \ - \ 3 \bar{\beta} \ + \ \bar{\tau}) \ - \ \Delta \bar{\epsilon} \ + \ \Delta \bar{\rho} \ - \ \bar{\delta} \bar{\alpha} \ - \ 2 \bar{\delta} \bar{\pi}] \nonumber \\[2mm]
&&  + \ \bar{\Psi}_4[\bar{\kappa}(\gamma \ - \ \bar{\gamma} \ - \ 3 \bar{\mu}) \ + \ \rho(4 \bar{\beta} \ - \ \bar{\tau}) \ + \ \bar{\rho}(\alpha \ - \ 3 \bar{\beta} \ + \ \bar{\tau}) \ + \ 4 \bar{\epsilon}(3 \bar{\beta} \ - \ \bar{\tau} \ - \ \alpha) \ -\ \bar{\sigma} \tau \nonumber \\[2mm]
&&  - \ \Delta \bar{\kappa} \ + \ 4 \bar{\delta} \bar{\epsilon} \ - \ \bar{\delta}\bar{\rho}]\, ,
\end{eqnarray}

\begin{eqnarray}
B_{mm}^{(1)} &=& \Delta \Delta \Psi_0 \ - \ \Delta \delta \Psi_1 \ - \ \delta \Delta \Psi_1 \ + \ \delta \delta \Psi_2 \nonumber \\[2mm]
&&  + \ (2 \mu \ - \ 7 \gamma \ + \ \bar{\gamma}) \Delta \Psi_0 \ + \ \nu \delta \Psi_0 \ - \ \bar{\nu} \bar{\delta} \Psi_0 \nonumber \\[2mm]
&&  + \ \bar{\nu} \Psi_1 \ + \ (7 \tau \ - \ \bar{\alpha} \ + \ 3 \beta) \Delta \Psi_1 \ + \ (5 \gamma \ -  \ \bar{\gamma} \ - \ 3 \mu) \delta \Psi_1 \ + \ \bar{\lambda} \bar{\delta} \Psi_1 \nonumber \\[2mm]
&&  - \ \bar{\lambda} \mathrm{D} \Psi_2 \ - \ 5 \sigma \Delta \Psi_2 \ + \ (\bar{\alpha} \ - \ \beta \ - \ 6 \tau) \delta \Psi_2 \ + \ 4 \sigma \delta \Psi_3 \nonumber \\[2mm]
&&  + \ \Psi_0[\mu(\mu \ - \ 7 \gamma \ + \ \bar{\gamma}) \ + \ \nu(\bar{\alpha} \ - \ \beta \ - \ 3 \tau) \ + \ \bar{\nu}(4 \alpha \ - \ \pi) \ + \ 4 \gamma(3 \gamma \ - \  \bar{\gamma}) \nonumber \\[2mm]
&&  - \ \lambda \bar{\lambda} \ - \ 4 \Delta \gamma \ + \ \Delta \mu \ + \ \delta \nu] \nonumber \\[2mm]
&&  + \ 2 \Psi_1[2 \nu \sigma \ - \ \bar{\nu}(\epsilon \ + \ 2 \rho) \ + \ \bar{\lambda}(\pi \ - \ \alpha) \ + \ (\bar{\gamma} \ - \ 2 \gamma)(\beta \ + \ 2 \tau) \nonumber \\[2mm]
&&  + \ (\mu \ - \ \gamma)(5 \tau \ - \ \bar{\alpha} \ + \ 2 \beta)  \ + \  \Delta\beta \ + \  2\Delta\tau \ + \ \delta\gamma \ - \ \delta\mu ] \nonumber \\[2mm]
&&  + \ 3 \Psi_2[\kappa \bar{\nu} \ + \ \bar{\lambda} \rho \ + \ \sigma(3 \gamma \ - \ \bar{\gamma} \ - \ 3 \mu) \ + \ \tau(3 \tau \ - \ \bar{\alpha} \ + \ \beta) \ - \ \Delta \sigma \ - \ \delta \tau] \nonumber \\[2mm]
&&  + \ 2 \Psi_3[  - \ \kappa \bar{\lambda} \ + \ \sigma(\bar{\alpha} \ + \ \beta \ - \ 5 \tau) \ + \ \delta \sigma] \ + \ 2 \Psi_4 \sigma^2 \nonumber \\[2mm]
&&  + \ \mathrm{DD} \bar{\Psi}_4 \ - \ \mathrm{D} \delta \bar{\Psi}_3 \ - \ \delta \mathrm{D} \bar{\Psi}_3 \ + \ \delta \delta \bar{\Psi}_2 \nonumber \\[2mm]
&&  - \ 4 \bar{\lambda} \delta \bar{\Psi}_1 \ + \ 5 \bar{\lambda} \mathrm{D} \bar{\Psi}_2 \ + \ \sigma \Delta \bar{\Psi}_2 \ + \ (\bar{\alpha} \ - \ \beta \ + \ 6 \bar{\pi}) \delta \bar{\Psi}_2 \nonumber \\[2mm]
&&  + \ (\beta \ - \ 3 \bar{\alpha} \ - \ 7 \bar{\pi}) \mathrm{D} \bar{\Psi}_3 \ - \ \kappa \Delta \bar{\Psi}_3 \ + \ (\epsilon \ - \ 5 \bar{\epsilon} \ + \ 3 \bar{\rho}) \delta \bar{\Psi}_3 \ - \ \sigma \bar{\delta} \bar{\Psi}_3 \nonumber \\[2mm]
&&  + \ (7 \bar{\epsilon} \ - \ \epsilon \ - \ 2 \bar{\rho}) \mathrm{D} \bar{\Psi}_4 \ - \ \bar{\kappa} \delta \bar{\Psi}_4 \ + \ \kappa \bar{\delta} \bar{\Psi}_4 \nonumber \\[2mm]
&&  + \ 2 \bar{\Psi}_0 \bar{\lambda}^2 \ + \ 2 \bar{\Psi}_1[\bar{\lambda}(\bar{\alpha} \ + \ \beta \ - \ 5 \bar{\pi}) \ - \ \bar{\nu} \sigma \ - \ \delta \bar{\lambda}] \nonumber \\[2mm]
&&  + \ 3 \bar{\Psi}_2[\kappa \bar{\nu} \ + \ \bar{\lambda}(3 \bar{\epsilon} \ - \    \epsilon \ - \ 3 \bar{\rho}) \ + \ \bar{\mu} \sigma \ + \ \bar{\pi}(\bar{\alpha} \ - \ \beta \ + \ 3 \bar{\pi}) \ + \ \mathrm{D} \bar{\lambda} \ + \ \delta \bar{\pi}] \nonumber \\[2mm]
&&  + \ 2 \bar{\Psi}_3[2 \bar{\kappa} \bar{\lambda} \ - \ \kappa(2 \bar{\mu} \ + \ \bar{\gamma}) \ + \ \sigma(\bar{\tau} \ - \ \bar{\beta}) \ + \ (\bar{\rho} \ - \ \bar{\epsilon})(2 \bar{\alpha} \ - \ \beta \ + \ 5 \bar{\pi}) \nonumber \\[2mm]
&&  + \ (\epsilon \ - \ 2 \bar{\epsilon})(2 \bar{\pi} \ + \ \bar{\alpha}) \ - \     \mathrm{D} \bar{\alpha} \ - \ 2 \mathrm{D} \bar{\pi} \ - \ \delta \bar{\epsilon} \ + \ \delta \bar{\rho}] \nonumber \\[2mm]
&&  + \ \bar{\Psi}_4[\kappa(4 \bar{\beta} \ - \ \bar{\tau}) \ + \ \bar{\kappa}(\beta \ - \ \bar{\alpha} \ - \ 3 \bar{\pi}) \ + \ (\bar{\rho} \- \ 4 \bar{\epsilon})(\epsilon \ - \ 3 \bar{\epsilon} \ + \ \bar{\rho}) \ - \ \sigma \bar{\sigma} \nonumber \\[2mm]
&&  + \ 4 \mathrm{D} \bar{\epsilon} \ - \ \mathrm{D} \bar{\rho} \ - \ \delta \bar{\kappa}]\, ,
\end{eqnarray}

\begin{eqnarray}
\label{eq:BachNP11}
B_{nn}^{(1)} &=& \Delta \Delta \Psi_2 \ - \ \Delta \delta \Psi_3 \ - \ \delta \Delta \Psi_3 \ + \ \delta \delta \Psi_4 \nonumber \\[2mm]
&& - \ 4 \nu \Delta \Psi_1 \ + \ (\gamma \ + \ \bar{\gamma} \ + \ 6 \mu) \Delta \Psi_2 \ + \ 5 \nu \delta \Psi_2 \ - \ \bar{\nu} \bar{\delta} \Psi_2 \nonumber \\[2mm]
&& + \ \bar{\nu} \mathrm{D} \Psi_3 \  + \ (3 \tau \ - \ \bar{\alpha} \ - \ 5 \beta) \Delta \Psi_3 \ - \ (3 \gamma \ + \   \bar{\gamma} \ + \ 7 \mu) \delta \Psi_3 \ + \ \bar{\lambda} \bar{\delta} \Psi_3 \nonumber \\[2mm]
&& - \ \bar{\lambda} \mathrm{D} \Psi_4 \ - \ \sigma \Delta \Psi_4 \ + \ (\bar{\alpha} \ + \ 7 \beta \ - \ 2 \tau) \delta \Psi_4 \nonumber \\[2mm]
&& + \ 2 \Psi_0 \nu^2 \ + \ 2 \Psi_1[\nu(\gamma \ - \ \bar{\gamma} \ - \ 5 \mu) \ + \ \lambda \bar{\nu} \ - \ \Delta \nu] \nonumber \\[2mm]
&& + \ 3 \Psi_2[\mu(\gamma \ + \ \bar{\gamma} \  + \ 3 \mu) \ + \ \nu(\bar{\alpha} \ + \ 3 \beta \ - \ 3 \tau) \ - \ \lambda \bar{\lambda} \ - \ \bar{\nu} \pi \ + \ \Delta \mu \ + \ \delta \nu] \nonumber \\[2mm]
&& + \ 2  \Psi_3[\bar{\nu}(\epsilon \ - \ \rho) \ + \ \bar{\lambda}(\alpha \ + \ 2 \pi) \ + \ \gamma(2 \tau \ - \ \bar{\alpha} \ - \ 4 \beta) \ + \ \bar{\gamma}(\tau \ - \ \beta) \nonumber \\[2mm]
&& + \ \mu(5 \tau \ - \ 2 \bar{\alpha} \ - \ 9 \beta) \ + \ 2 \nu \sigma \ - \ \Delta \beta \ + \ \Delta \tau \ - \ \delta \gamma \ - \ 2 \delta \mu] \nonumber \\[2mm]
&& + \ \Psi_4[\kappa \bar{\nu} \ + \ \bar{\lambda}(\rho \ - \ 4 \epsilon)  \ - \ \sigma(\gamma \ + \ \bar{\gamma} \ + \ 3 \mu) \ + \ 4 \beta(3 \beta \ + \ \bar{\alpha}) \ + \ \tau(\tau \ - \ \bar{\alpha} \ - \ 7 \beta) \nonumber \\[2mm]
&& - \ \Delta \sigma \ + \ 4 \delta \beta \ - \ \delta \tau] \ + \ \mbox{c.c.}\, .
\end{eqnarray}\\
The second Ricci tensor part $B^{(2)}_{\mu\nu}=\frac{1}{2}\tensor{R}{^\alpha^\beta} \ \! \tensor{C}
{_\mu_\alpha_\nu_\beta}$ is then given by~\cite{Sv}\\
\begin{eqnarray} 
   && B^{(2)}_{ll} \ = \ \Phi_{20}\Psi_0 \ + \ \Phi_{02}\bar{\Psi}_0-2\Phi_{01}\Psi_1 \ - \  2\Phi_{10}\bar{\Psi}_1 \ + \ \Phi_{00}(\Psi_2 \ + \ \bar{\Psi}_2)\, , \nonumber \\[2mm]
&&
    B^{(2)}_{ln} \ = \ \Phi_{21}\Psi_1 \ + \ \Phi_{12}\bar{\Psi}_1 \ + \ \Phi_{01}\Psi_3 \ + \ 2\Phi_{10}\bar{\Psi}_3 \ - \ 2\Phi_{11}(\Psi_2 \ + \ \bar{\Psi}_2)\, ,\nonumber \\[2mm]
&&
    B^{(2)}_{lm} \ = \ \Phi_{21}\Psi_0 \ - \ 2\Phi_{11}\Psi_1 \ + \ \Phi_{01}(\Psi_2 \ - \ 2\bar{\Psi}_2) \ + \ \Phi_{02}\bar{\Psi}_1 \ + \ \Phi_{00}\bar{\Psi}_3\, , \nonumber \\[2mm]
&&
    B^{(2)}_{nm} \ = \ \Phi_{22}\Psi_1 \ + \ \Phi_{12}(\bar{\Psi}_2-2\Psi_2) \ + \ \Phi_{02}\Psi_3 \ - \ 2\Phi_{11}\bar{\Psi}_3 \ + \ \Phi_{01}\bar{\Psi}_4\, , \nonumber \\[2mm]
&&
    B^{(2)}_{mm} \ = \ \Phi_{22}\Psi_0 \ - \ 2\Phi_{12}\Psi_1 \ + \ \Phi_{02}(\Psi_2 \ + \ \bar{\Psi}_2) \ - \ \Phi_{01}\bar{\Psi}_3 \ + \ \Phi_{00}\bar{\Psi}_4\, , \nonumber \\[2mm]
&&
    B^{(2)}_{nn} \ = \ \Phi_{22}(\Psi_2 \ + \ \bar{\Psi}_2) \ - \ 2\Phi_{12}(\Psi_3 \ + \ \bar{\Psi}_3) \ + \ \Phi_{02}\Psi_4 \ + \ \Phi_{20}\bar{\Psi}_4\, .
    \label{B7cc}
\end{eqnarray}\\
\end{widetext}
The rest of the equations can be obtained by complex conjugation. The tracelessness
of $B_{\mu \nu}$ implies that $B_{ln} = B_{m\bar{m}}$.
Despite the complicated nature of the above expressions, they are drastically simplified under the assumption of spherical symmetry. 

Particularly relevant in this paper are Petrov ``type-D'' spacetimes. The latter correspond
to gravitational fields of isolated massive objects that far enough look like point sources. {\cdblue Although we do not know whether the Goldberg--Sachs theorem and its corollary~\cite{Chandra}, which automatically ensures $\kappa = \sigma =
\nu = \lambda = 0$ for Petrov type D solutions, also holds in WCG, $\kappa = \sigma =
\nu = \lambda = 0$ can be often easily achieved for spacetimes possessing some kind of symmetry.}

In Secs.~\ref{MK}, \ref{Sec.III.C.cv} and \ref{Sec.V.cd}, we have seen that all presented solutions  are of type D.
For Petrov ``type D'' spacetimes (\ref{eq:BachNP00})-(\ref{B7cc}) reduce {\cdblue under the condition $\kappa = \sigma =
\nu = \lambda = 0$} to\\
\begin{widetext}
\begin{eqnarray}
    && B_{ll} \ = \ \mathrm{D}\mathrm{D}\Psi_2 \ - \ (\epsilon \ + \    \bar{\epsilon} \ + \ 6\rho)\mathrm{D}\Psi_2 \ + \ 3\Psi_2\left[\rho(\epsilon \ + \  \bar{\epsilon} \ + \ 3\rho) \ - \ \mathrm{D}\rho\right] \ + \ \Phi_{00}\Psi_2 \ + \ \mbox{c.c.}\, , \nonumber \\[3mm]
&&
    B_{ln} \ = \ -\ \mathrm{D}\Delta\Psi_2 \ - \ \bar{\delta}\delta\Psi_2 \ + \ (\bar{\mu} \ - \ 3\mu)\mathrm{D}\Psi_2 \ + \ (2\rho \ - \ \epsilon \ - \ \bar{\epsilon})\Delta\Psi_2 \ + \ (\alpha \ - \     \bar{\beta}\ -\ 2\pi)\delta\Psi_2  \nonumber \\[2mm] 
    &&\mbox{\hspace{12mm}}  + \ (\bar{\pi} \ + \ 3\tau)\bar{\delta}\Psi_2 \ + \ 3\Psi_2\left[\mu(2\rho \ - \ \epsilon \ - \   \bar{\epsilon}) \ - \ \bar{\mu}\rho \ + \ \pi\bar{\pi} \ + \ \tau(2\pi \ - \ \alpha \ + \ \bar{\beta}) \ - \ \mathrm{D}\mu \ + \ \bar{\delta}\tau\right] \nonumber \\[2mm] 
    &&\mbox{\hspace{12mm}}- \ 2\Phi_{11}\Psi_2 \ + \ \mbox{c.c.}\, ,  \nonumber \\[3mm]
&&
    B_{lm} \ = \ \mathrm{D}\delta\Psi_2 \ - \ (\bar{\pi}+3\tau)\mathrm{D}\Psi_2 \ -\  (\epsilon \ - \ \bar{\epsilon} \ + \ 3\rho)\delta\Psi_2 \nonumber \\[2mm] 
    &&\mbox{\hspace{12mm}} + \  3\Psi_2\left[\bar{\pi}\rho \ + \ \tau(\epsilon \ - \ \bar{\epsilon} \ + \ 3\rho) \ - \ \mathrm{D\tau}\right] \ - \ \mathrm{D}\delta\bar{\Psi}_2\nonumber \\[2mm]
    &&\mbox{\hspace{12mm}}- \ \delta\mathrm{D}\bar{\Psi}_2 \ + \ (\alpha \ + \   \beta \ - \ 5 \bar{\pi})\mathrm{D}\Psi_2 \ + \ (\epsilon \ - \    \bar{\epsilon} \ + \ 5\bar{\rho})\delta\bar{\Psi}_2 \nonumber \\[2mm] 
    &&\mbox{\hspace{12mm}}+ \ 3 \bar{\Psi}_2\left[\bar{\pi}(\epsilon \ - \ \bar{\epsilon}) \ + \ \bar{\rho}(4\bar{\pi} \ - \ \bar{\alpha} \ - \ \beta) \ - \ \mathrm{D}\bar{\pi} \ + \ \delta\bar{\rho}\right] \ + \ \Phi_{01}(\Psi_2 \ - \ 2\bar{\Psi}_2) \, , \nonumber \\[3mm]
&&
    B_{nm} \ = \ -\ \Delta\delta\Psi_2 \ - \ \delta\Delta\Psi_2 \ + \ (5\tau \ - \   \bar{\alpha} \ - \ \beta)\Delta\Psi_2 \ + \ (\gamma \ - \   \bar{\gamma} \ - \    5\mu)\delta\Psi_2 \nonumber \\[2mm] 
    &&\mbox{\hspace{12mm}} + \ 3\Psi_2\left[\mu(4\tau \ - \ \bar{\alpha-\beta}) \ + \    \tau(\bar{\gamma)} \ - \ \gamma) \ + \ \Delta\tau \ - \ \delta\mu\right] \ + \ (3\pi \ + \ \tau)\Delta\bar{\Psi}_2 \nonumber \\[2mm] 
    &&\mbox{\hspace{12mm}}+ \ 3\bar{\Psi}_2\left[\bar{\pi}(3\bar{\mu} \ - \ \gamma \ + \ \bar{\gamma}) \ + \ \bar{\mu}\tau \ + \ \Delta\bar{\pi}\right] \ + \ \Phi_{12}(\bar{\Psi}_2 \ - \ 2\Psi_2) \, , \nonumber \\[3mm]
    &&B_{mm} \ = \  \delta\delta\Psi_2 \ + \ (\bar{\alpha} \ - \ \beta \ - \ 6\tau)\delta\Psi_2 \ + \ 3\Psi_2\left[\tau (3\tau \ - \ \bar{\alpha} \ + \ \beta) \  - \ \delta\tau\right]\nonumber \\[2mm] 
    &&\mbox{\hspace{12mm}} + \ \delta\delta\bar{\Psi}_2 \ + \ (\bar{\alpha} \ - \ \beta \ + \ 6\bar{\pi})\delta\bar{\Psi}_2 \ + \ 3\bar{\Psi_2}\left[\bar{\pi}(\bar{\alpha} \ - \ \beta \ + \ 3\bar{\pi}) \ + \ \delta\bar{\pi}\right] \ + \  \Phi_{02}(\Psi_2 \ + \ \bar{\Psi}2)\, , \nonumber \\[3mm]
    &&B_{nn} \ = \  \Delta\Delta\Psi_2 \ + \ (\gamma \ + \ \bar{\gamma} \ + \ 6\mu)\Delta\Psi_2 \ + \ 3\Psi_2\left[\mu(\gamma \ + \ \bar{\gamma} \ + \ 3\mu) \ + \ \Delta\mu\right] \ + \ \Phi_{22}\Psi_2  \ + \  \mbox{c.c.}\, .
\end{eqnarray}\\
Even though the Bach equations are nonlinear in terms of the metric tensor, they are linear in the NP-Weyl scalars. \\
\end{widetext}

\end{document}